\documentclass[3p,onecolumn]{elsarticle}
\usepackage[utf8]{inputenc}
\usepackage[english]{babel}
\usepackage{amssymb}
\usepackage{amsmath}
\usepackage{amsfonts}
\usepackage{url}
\usepackage{hyperref}
\usepackage{physics}
\usepackage{bm}
\usepackage{graphicx}
\usepackage{color}
\usepackage{subcaption}
\usepackage{tikz}
\usepackage{pgfplots}
\usepackage{multirow}

\pgfplotsset{compat=1.16}

\usepackage[normalem]{ulem}

\graphicspath{ {./img/} }

\newcommand{\E}[1]{\cdot 10^{#1}}

\newcommand{\Bi}{\mathrm{Bi}}
\newcommand{\mrs}{\mathrm{s}}
\newcommand{\mre}{\mathrm{e}}
\newcommand{\mrn}{\mathrm{n}}
\newcommand{\mrp}{\mathrm{p}}
\newcommand{\typ}{\mathrm{typ}}
\newcommand{\degC}{^\circ \mathrm{C}}
\DeclareMathOperator{\arcsinh}{arcsinh}

\usepackage{todonotes}

\title{Systematic derivation and validation of a reduced thermal-electrochemical model for lithium-ion batteries using asymptotic methods}
\author[1,2]{Ferran Brosa Planella\corref{cor}}
\ead{Ferran.Brosa-Planella@warwick.ac.uk}
\author[1]{Muhammad Sheikh}
\author[1,2]{W. Dhammika Widanage}

\address[1]{WMG, University of Warwick, Gibbet Hill Road, Coventry, CV4 7AL, United Kingdom}
\address[2]{The Faraday Institution, Harwell Campus, Didcot, OX11 0RA, United Kingdom}

\cortext[cor]{Corresponding author: Ferran.Brosa-Planella@warwick.ac.uk}

\begin{document}

\begin{abstract}
The widely used Doyler-Fuller-Newman (DFN) model for lithium-ion batteries is too computationally expensive for certain applications, which has motivated the appearance of a plethora of simpler models. These models are usually posed in an \emph{ad hoc} manner, leading to inconsistencies with the DFN model and to multiple formulations of the same model, with the Single Particle Model (SPM) being a very good example of the latter. In this work, we discuss the concept of SPM-type models showing that, despite the multiple formulations found in the literature, these models always follow the same structure, and we extend this discussion to models accounting for thermal effects. Then, we present a Thermal Single Particle Model with electrolyte (TSPMe) derived in a systematic manner using asymptotic techniques. The validation of the TSPMe against a thermal DFN model shows very high accuracy with a computational cost over forty times smaller. The comparison against experimental data shows that the model does a reasonable job predicting the behaviour of a real battery, but a very good parameter set is required to obtain accurate predictions.
\end{abstract}


\maketitle

\section{Introduction}

With the electrification of vehicles and the spread of portable electronic devices, lithium-ion batteries have become a key technology for energy storage. This has motivated a quest to design batteries that can store more energy, last longer, and operate efficiently and safely. Mathematical models are an invaluable tool for battery design and control as they provide a cheap, safe and fast alternative to experiments. 

There exist multiple approaches to deterministic models for lithium-ion batteries, and they can usually be classified into two broad categories: equivalent-circuit models and physics-based models. Equivalent-circuit models describe the battery by assuming it has the structure of a particular electrical circuit. The morphology of this circuit is a modelling assumption, and the values of the different parameters need to be determined from fitting the model to experimental data (see \cite{Hu2012,Widanage2016} for examples). Equivalent-circuit models are widely used in battery management systems (BMS) because they are computationally cheap to simulate \cite{Farmann2016}. However, they offer little insight on the battery internal states and therefore it is very complicated to extend these models to account for additional physics. A detailed description of equivalent-circuit models can be found in \cite{Plett2015}.

In this work, we focus on physics-based models. These models are derived from fundamental physical laws and have the advantage that they provide significant insight on the internal states of the battery. However, due to their complexity, they are computationally more expensive than equivalent-circuit models. The most widely used physics-based model is the Doyle-Fuller-Newman (DFN) model, originally posed in \cite{Fuller1994,Fuller1994a}. This model accounts for conservation of mass and charge in both electrode and electrolyte, and the four equations are coupled together through the reaction current density term that describes the electrode-electrolyte intercalation reaction. One of the key aspects for the success of the DFN model is that it assumes a simplified geometry that captures the main features of a battery but is much more affordable to solve than trying to resolve all the phenomena at a microscale level \cite{Lu2020}. The geometry of the DFN model assumes that most of the variables (electrolyte concentration and potentials both in electrode and electrolyte) can be resolved at a macroscale level, in which the material is assumed to be homogeneous. In order to account for the effects of the porous material on the transport equations in the homogeneous material, effective transport parameters are used (often using the Bruggeman correlation \cite{Tjaden2016}). The transport of intercalated lithium is described by assuming that at each point of the electrode there is a representative spherical particle in which lithium diffuses. Even though it is not explicitly derived in \cite{Fuller1994}, this is a homogenised model and it can be formalised from the mathematical point of view (see \cite{Hunt2020} for details). A more detailed description of the DFN model can be found in the handbooks \cite{Plett2015,Newman2004}.

The DFN model is still quite complex as it involves a coupled system of differential-algebraic equations (DAEs). Hence, it is not suitable for applications where speed is crucial, such as battery optimization and control, so simpler models have been posed in the literature. One of the best-known examples is the Single Particle Model (SPM). The key idea of this model is that we can use one single particle in each electrode to represent the behaviour of all particles. SPMs may \cite{Moura2017} or may not \cite{Bizeray2018} include electrolyte dynamics. In both cases the model is much simpler than the DFN model as most of the variables of interest can be easily computed from the concentrations in the electrode particles and the electrolyte.

The SPMs are widely used, but in their origin there are still two main challenges. First, most of these models are posed on an \emph{ad hoc} basis and therefore the connection to the DFN (or similar) model is lost. This is far from ideal because, in the process of derivation, inconsistencies might be introduced and certain features lost. In addition, it makes it hard to add new physics in a consistent way. The second issue, closely related to the first, is that we can find multiple Single Particle Models in the literature following different formulations \cite{Moura2017,Bizeray2018,Kemper2013,Perez2016,Marquis2019,Richardson2020}. Over the last few years, there has been a surge in the asymptotic analysis of battery models \cite{Marquis2019,Richardson2020,Marquis2020,Richardson2012,Moyles2019asymptotic,Moyles2018experimental,Hennessy2019} with the goal of obtaining reduced models in a systematic manner. The different authors take different assumptions to reduce the models. One big difference is that \cite{Richardson2012,Moyles2019asymptotic,Moyles2018experimental,Hennessy2019} assume fast diffusion in the electrodes and thus do not reduce to SPM-type models, while \cite{Marquis2019,Richardson2020,Marquis2020} retain this feature. Given that our interest on this work is on SPM-type models, we will focus our attention on the latter.

The key idea behind asymptotic methods is to analyse the dimensionless groupings that appear in the model, determine which can be assumed to be small and perform an expansion in the limit where these parameters tend to zero to obtain a simpler model. More details on these methods can be found in the handbook \cite{Hinch1991}. These methods present multiple advantages. First, they are systematic so they can be applied to all the models, even the very complex ones. Second, they also ensure the consistency between the full and reduced models, and thus the reduced model satisfies the same physical laws as the full one (e.g. conservation of mass). And third, we can validate the assumptions taken and estimate the error of the reduced model \emph{a priori} (i.e. before even solving the model).

So far we referred to electrochemical models only, but a similar description holds for coupled thermal-electrochemical models. These are usually posed as a coupling between the DFN model, to describe the electrochemistry, and some sort of thermal model \cite{Plett2015,Gu2000,Srinivasan2003}. Due to their complexity, simpler models have been posed \emph{ad hoc}, usually based on SPM-type electrochemical models \cite{Perez2016}. The asymptotic analysis of thermal-electrochemical models is even more recent as it is a natural evolution of the asymptotic reduction of pure electrochemical models. To our knowledge, the only references in the literature on asymptotic reductions of thermal-electrochemical models are \cite{Hennessy2019,Timms2020}. In \cite{Hennessy2019} a multilayer cell is considered but with the electrode fast diffusion electrochemical model presented in \cite{Moyles2019asymptotic}; while in \cite{Timms2020} a model for single-layer pouch cells based on DFN and the SPMe of \cite{Marquis2019}.

The goal of this paper is to derive a SPM-type thermal-electrochemical model for a multi-layer cell, using an asymptotic method based on minimal assumptions. The resultant model, which we refer to as Thermal Single Particle Model with electrolyte (TSPMe), is validated against the full thermal DFN model and experimental data to show its validity and proof that it is suitable for practical applications. To the best of our knowledge, this model is the first instance of an asymptotically derived thermal-electrochemical model for a multi-layer battery that also accounts for finite diffusion in the electrode particles.

\section{Discussion of Doyler-Fuller-Newman model and Single Particle Models}\label{sec:SPM_discussion}

Before presenting the formulation for the Thermal Single Particle Model with electrolyte, we discuss the nature of Single Particle Models and compare it with that of the DFN model. Some of the most popular reduced models for lithium-ion batteries are Single Particle Models (SPM), and in the literature we can find several different models referred to as such \cite{Moura2017,Bizeray2018,Kemper2013,Perez2016,Marquis2019,Richardson2020}. These are usually based on slightly different sets of assumptions and therefore their formulations are different as well. Because the different models are referred to by the same name, it makes the term ``Single Particle Model'' very confusing. With the hope of shedding some light into this issue, in this section we discuss what are the fundamental features that constitute a Single Particle Model, understanding it as a type of model rather than a specific formulation. We refer to models of this kind as SPM-type models, which applies to both models with and without electrolyte dynamics. However, given that the model without electrolyte dynamics (SPM) is a particular case to the model with electrolyte dynamics (SPMe), we consider the latter in this analysis. By considering what are the essential features of an SPM-type model we can better understand and classify existing models in the literature.

\subsection{Electrochemical models}\label{sec:EC_models}
\begin{figure}
    \centering
    \begin{subfigure}{0.49\textwidth}
      \centering
      \includegraphics[scale=1]{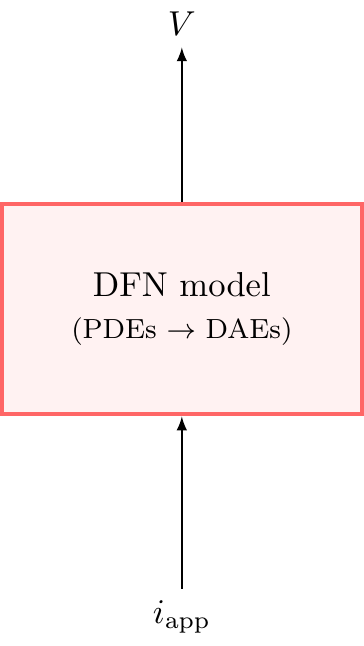}  
      \caption{DFN model}
    \end{subfigure}\hfill
    \begin{subfigure}{0.49\textwidth}
      \centering
      \includegraphics[scale=1]{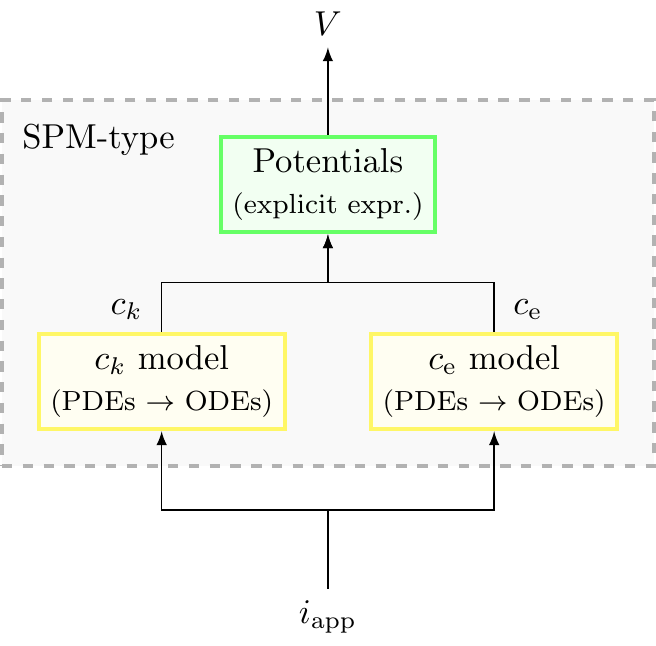}  
      \caption{SPM-type}
    \end{subfigure}
    \caption{Block diagram of standard electrochemical models: (a) the Doyler-Fuller-Newman model (DFN) and (b) the Single Particle Model type (SPM-type). Each block is a model (or part of a model) that converts the input variable into the output variable (or variables). In brackets, we define the the type of expressions that compose the model (e.g. PDEs, ODEs, explicit expressions) and, if applicable, the type of expressions after spatial discretisation (e.g. DAEs or ODEs). The colour of the blocks indicates the complexity of the model: red for a system of differential-algebraic equations (DAEs, high complexity), yellow for a system of ordinary differential equations (ODEs, medium complexity) and green for explicit expressions (low complexity).}
    \label{fig:block_diagram_EC}
\end{figure}

We start focusing on purely electrochemical models. By electrochemical models we refer to models that only account for mass and charge transport (in both the electrodes and the electrolyte) but do not include further physical effects such as thermal, mechanical or degradation. The classic example of an electrochemical model is the Doyle-Fuller-Newman (DFN) model \cite{Fuller1994}. To better understand the electrochemical models, we can represent them as a block diagram, as shown in Figure~\ref{fig:block_diagram_EC}. In these diagrams, we have an input (or inputs) that are known and want to calculate an output (or outputs) which are unknown. To convert an input into an output we use a model, represented in the diagram by a block. Note that this model could be anything that converts the input into an output, but here we consider it from the mathematical point of view, in which the model is composed of a set of differential and/or algebraic equations that need to be solved. The most typical setup for electrochemical models is to use the applied current density ($i_\mathrm{app}$) as input and calculate the terminal voltage ($V$) as an output. Other setups, such as inputting the voltage are common as well, but make the reduced models significantly more expensive as we discuss later. In addition, sometimes there are other variables that might be of interest and hence be part of the model output. 

We start by considering the DFN model \cite{Fuller1994}. The model is composed of a system of parabolic and elliptic partial differential equations (PDEs), which are represented as a single block in the diagram indicating the need to solve the full model at every step output is required (see Figure~\ref{fig:block_diagram_EC}a). A very common method to solve partial differential equations is to use the method of lines \cite{Schiesser2016}, which consists of discretising the system in space, for example using finite volume methods \cite{LeVeque2002} or finite element methods \cite{Johnson2009}, and leaving the time derivative without discretising. This method transforms parabolic PDEs into ordinary differential equtions (ODEs) and elliptic PDEs into algebraic equations. Therefore, for the DFN model, which is a coupled system of parabolic and elliptic PDEs, we obtain after discretising a system of differential-algebraic equations (DAEs). Solving numerically a system of DAEs is a complex task, hence the model is very expensive to compute.

The key aspect of SPM-type models is that they break this block into simpler blocks (see Figure~\ref{fig:block_diagram_EC}b), and we can group these blocks into two main steps. The first step involves calculating from the input current the concentrations in the representative electrode particles ($c_k$ with $k \in \{\mrn,\mrp\}$ for negative and positive electrode, respectively), and in the electrolyte ($c_\mre$, if not neglected). All the other variables of interest can then be calculated from these concentrations using explicit expressions, or even closed-form expressions in some cases. The first step in the SPM-type models (``$c_k$ model'' and ``$c_\mre$ model'' blocks in Figure~\ref{fig:block_diagram_EC}b) involves solving up to three decoupled partial differential equations: two PDEs for the concentration of lithium in the particles (one for each electrode) and one PDE for the ion concentration in the electrolyte. If the electrolyte dynamics are not considered we only need to solve the two PDEs for the particles. Because the PDEs are decoupled, we can solve them independently which results in a simple problem. To numerically solve these PDEs we can use the method of lines again, in order to obtain a system of ODEs which can be solved using standard methods. These systems of ODEs are much simpler to solve than the systems of DAEs that arise from the DFN model. Additionally, further techniques can be used to speed up the calculations of the solution of the ODE system (e.g residue grouping \cite{Smith2008} or balanced truncation \cite{Jun2015}). The second step (``Potentials'' block in Figure~\ref{fig:block_diagram_EC}b) consists of evaluating the explicit expression for the terminal voltage and for any other variable of interest (such as currents or potentials) which takes the electrolyte and electrode concentrations as inputs. Because this step involves evaluating an explicit expression rather than solving an equation, it is computationally cheap. In addition, we can limit this step to evaluating only the variables we are interested in. The main advantage of this approach is that it fully decouples the problem, so each concentration is the solution of an independent PDE and therefore much cheaper to solve than the DFN model. Note, however, that this decoupling only occurs in the case where current is the input of the model. If instead we want to simulate a potentiostatic discharge, we need to impose an algebraic constraint on the voltage and thus we need to solve a system of DAEs. This system, even though it is more complex than the standard decoupled SPM-type models, is still much simpler than the DFN model.

In the literature, we find several instances of SPM-type models, posed either in an \emph{ad hoc} manner \cite{Moura2017,Bizeray2018,Kemper2013,Perez2016} or a systematic manner \cite{Marquis2019,Richardson2020}. In most cases both approaches yield very similar results, but asymptotic methods are systematic and, therefore, more reliable. Moreover, the systematic approach allows us to easily extend the models to account for extra physical phenomena in a rigorous way, ensuring that underlying physical principles of the DFN model still hold for the SPM-type models. These systematically derived models are obtained by applying asymptotic techniques, in which the size of the dimensionless groupings appearing in the model is used to decide which terms of the equations can be neglected. There are different sets of assumptions that lead to the derivation of SPM-type models (see \cite{Marquis2019,Richardson2020} for two different examples) but they yield very similar results \cite{Timms2020Corrigendum}. However, all of these assumptions are based on the same key idea that allows for SPM-type models: the battery operates in a regime where deviations from the equilibrium potential are small, which means that the battery operates at moderate to low C-rates. At high C-rates the discrepancies between DFN and SPM-type models are considerable, mainly due to the effect of the electrolyte, and thus the DFN model should be used \cite{Moura2017,Marquis2019}. Note that the definition of what is considered a ``high C-rate" for the applicability of these models is not an absolute definition, but it depends on the model parameters.

In the scenario of moderate to low C-rates, the analysis is as follows. If the deviations from equilibrium potential are small, then all the particles within an electrode are approximately at the same potential, which is around the equilibrium (or open-circuit) potential, and therefore they behave very similarly. In consequence, we can define a representative particle (or average particle) for each electrode, and then solve the lithium diffusion problem only for those two particles. Because all the particles behave similarly, the reaction current density for the lithium intercalation must be identical for all particles within an electrode and thus, using a conservation of charge argument, we can calculate it directly from the input current. This is a crucial result because the reaction current density was the condition that coupled the four conservation equations in the DFN model (conservation of mass and charge in both electrode and electrolyte). If the reaction current density is known \emph{a priori}, each of the four problems can now be treated separately and, moreover, the potential equations both in the electrodes and the electrolyte can be integrated analytically. Therefore, the only equations to solve numerically are the ones governing the concentrations and everything else can be calculated from explicit expressions, as shown in Figure~\ref{fig:block_diagram_EC}b. This reasoning can be formalised into a rigorous mathematical analysis using asymptotic techniques, as shown in \ref{sec:reduction_EC_model}. If additional assumptions are taken, there might be further simplifications that take place in the model which may yield different expressions for the potentials (``Potentials'' block in Figure~\ref{fig:block_diagram_EC}b). These further simplifications are discussed in Section \ref{sec:further_simplifications}.

\subsection{Thermal-electrochemical models}\label{sec:TEC_models}
\begin{figure}
    \centering
    \begin{subfigure}{0.48\textwidth}
      \centering
      \includegraphics[scale=1]{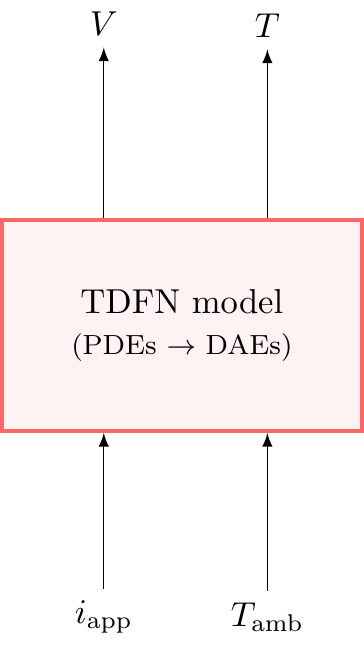}  
      \caption{TDFN model}
    \end{subfigure}\hfill
    \begin{subfigure}{0.48\textwidth}
      \centering
      \includegraphics[scale=1]{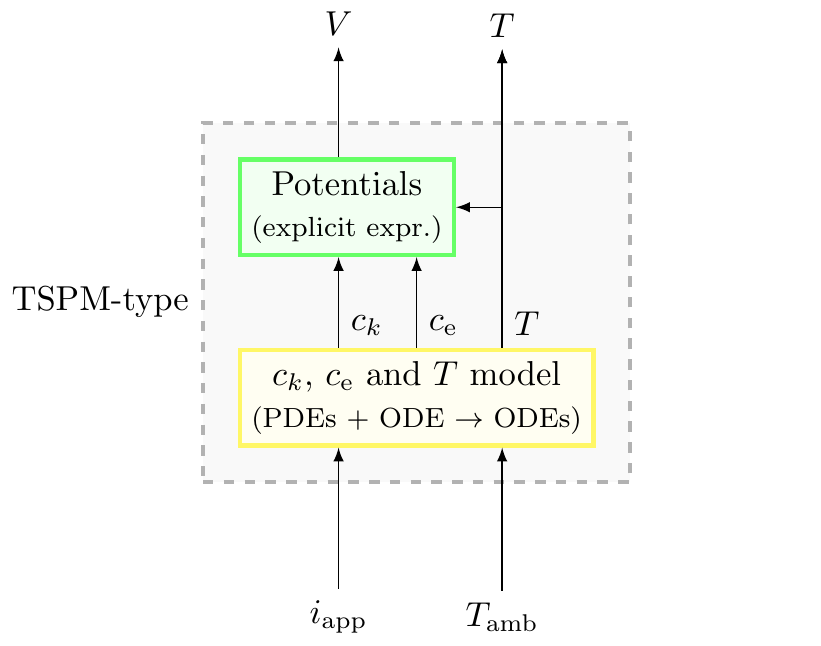}  
      \caption{TSPM-type}
    \end{subfigure}
    \caption{Block diagram of thermal-electrochemical models: (a) the thermal Doyler-Fuller-Newman model (TDFN) and (b) the Thermal Single Particle Model type (SPM-type). Each block is a model (or part of a model) that converts the input variable into the output variable (or variables). In brackets, we define the the type of expressions that compose the model (e.g. PDEs, ODEs, explicit expressions) and, if applicable, the type of expressions after spatial discretisation (e.g. DAEs or ODEs). The colour of the blocks indicates the complexity of the model: red for a system of differential-algebraic equations (DAEs, high complexity), yellow for a system of ordinary differential equations (ODEs, medium complexity) and green for explicit expressions (low complexity).}
    \label{fig:block_diagram_TEC}
\end{figure}

After considering electrochemical models, we focus on the coupled thermal-electrochemical models. These models, apart from the physics discussed in the previous section, describe how the temperature evolves in the battery. The two models are strongly coupled because the parameters of the electrochemical model depend on temperature and the heat generated in the battery depends on the electrochemical processes. This coupling adds an extra layer of complexity to thermal-electrochemical models compared to pure electrochemical models. We can visualise this extra layer of complexity in the form of a block diagram, as shown in Figure~\ref{fig:block_diagram_TEC}. Compared to the electrochemical models in Figure~\ref{fig:block_diagram_EC}, the thermal-electrochemical model has additional blocks to process the thermal inputs and outputs. In particular, the inputs for thermal-electrochemical models are applied current density and ambient temperature ($T_\mathrm{amb}$), and the new outputs are voltage and cell temperature ($T$).

The classic example of a thermal-electrochemical model would be the DFN coupled with a thermal model as shown in \cite{Plett2015,Gu2000,Srinivasan2003}. This model is even more computationally expensive than the standard DFN model, and another issue arises when considering a battery composed of multiple layers (we refer to each layer as a ``cell''). As discussed in \cite{Hunt2020}, in this situation the electrochemical behaviour needs to be described at a cell level, but the thermal behaviour needs to be described at a battery level (i.e. across multiple cells). This brings up a multi-scale coupling in the model that requires a hierarchy of submodels, similar to the one between electrodes and particles in the DFN model.

\begin{figure}
    \centering
    \includegraphics[scale=1]{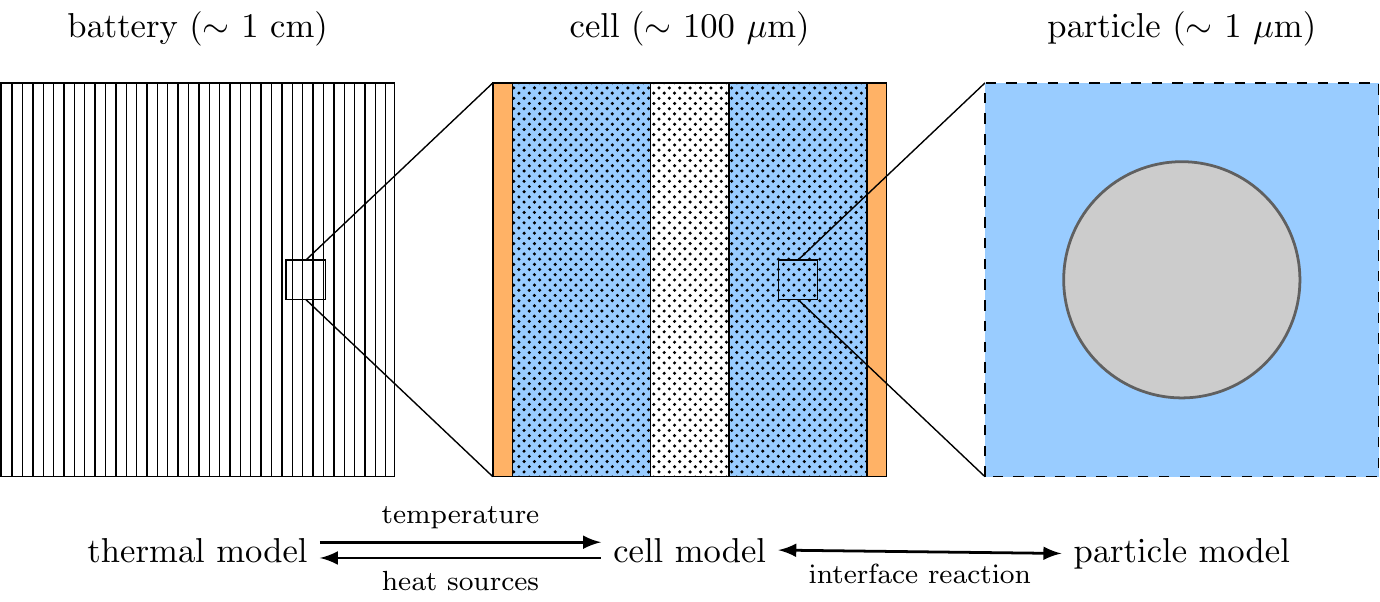}
    \caption{Sketch of the multiple scales involved in a thermal-electrochemical model. The thermal model is posed at the battery level, but it is coupled to the cell level through temperature dependence of parameters and heat generation terms. The cell model, in turn, is coupled with the particle model through the interface reaction.}
    \label{fig:multiscale}
\end{figure}

The thermal model is posed at the battery level because heat can flow from one cell to the neighbouring ones. However, ions cannot flow from one cell to the neighbouring ones, so the electrochemical models should be posed independently for each cell. Even in the case for cylindrical and prismatic cells, composed of big electrodes rolled up around a mandrel, the path for ions to go from one layer to the neighbouring ones is so long compared to the electrode thickness that these effects can be neglected. The two problems are strongly coupled, because the electrochemical properties of each cell can depend on the temperature while the heat generated by the electrochemistry impacts the evolution of temperature. The coupling between these two problems at different scales imposes a hierarchy similar to the classic DFN model. There, we solve for the electrode potential and the electrolyte concentration and potential across the cell, and at each point of the electrode domain there is a representative particle that describes the intercalated lithium concentration at the microscale. Here, we solve for temperature across the battery and at each point there is a representative cell that describes the electrochemical behaviour. Moreover, if the chosen electrochemical model is the DFN, then the hierarchy is at three levels: battery, cell and particle. These relations are shown in Figure~\ref{fig:multiscale}.

The thermal-electrochemical coupling at multiple scales results, after spatial discretisation, into a system of DAEs again, even when using an SPM-type electrochemical model, because for each point at the battery level we have a cell to represent the electrochemical behaviour. These cells are connected in parallel and therefore the total applied current is divided over the different cells, but not in an even way as it depends on the cell temperature. Therefore, we need additional algebraic constraints for the terminal voltage and applied current of each cell. Algebraic constraints were one of the main causes of the complexity of the DFN model, therefore, in this case too, we look for methods to simplify the model. The starting point is the DFN model coupled to a thermal model (see \cite{Plett2015,Gu2000,Srinivasan2003} for details). The simplifications on the electrochemical part of the model follow the procedure explained in Section \ref{sec:EC_models}, so here we focus on the thermal part only. In order to get rid of the DAEs again, we need to work in a regime in which the different instances of the electrochemical model (i.e. the different cells) can be considered to be very similar. It is the same idea as taking that all the particles work similarly within each electrode in the SPM model. There are two main assumptions that lead to this decoupling: the assumption that the temperature variation is small enough to not cause significant variations in the electrochemical parameters, and the assumption that the temperature is approximately homogeneous within the battery. The validity of each assumption depends on the battery chemistry, geometry and cooling conditions. In this work, we focus on the latter assumption as it is reasonable for the batteries and the conditions we are interested in: LG M50 cylindrical batteries cycled in a thermal chamber without any active cooling. The physical reasoning behind the reduction of the thermal model is that in this situation the bottleneck in the heat transfer is the heat dissipation to the environment rather than the heat transport inside the battery. Then, the thermal gradients inside the battery are negligible and the temperature is closely captured by the average temperature and we find that each cell of the battery behaves very similarly so we only need to solve one instance of the electrochemical model. The details for the derivation of the reduced thermal model are shown in \ref{sec:derivation_thermal}.

In terms of the block diagram, the TSPMe works in the following way. A system of coupled PDEs and ODEs (the latter for the temperature) is solved from the input variables, which in this case are the input current and ambient temperature. All the other quantities of interest can then be calculated from them. The main difference compared to the purely electrochemical models is that, if the parameters depend on temperature, the system of PDEs and ODEs is now coupled so concentrations and temperature need to be solved simultaneously (see Figure~\ref{fig:block_diagram_TEC}).

\section{Thermal Single Particle Model with electrolyte (TSPMe)}\label{sec:TSPMe}
Having discussed the background and interpretation of both DFN and SPM-type models, we can now present the Thermal Single Particle Model with electrolyte (TSPMe). The details on the full derivation from the DFN model are given in \ref{sec:derivation_TSPMe}. The TSPMe is a reduced thermal-electrochemical model for lithium-ion batteries. It consists of an SPMe for the electrochemistry coupled with a lumped thermal model for the cell, which requires solving three diffusion equations for the concentrations (two for a representative particle in each electrode and one for the electrolyte) and a first-order ODE for the average cell temperature. This model can be further simplified for some particular cases, as presented in Section \ref{sec:further_simplifications}.

\begin{figure}
    \centering
    \includegraphics[scale=1]{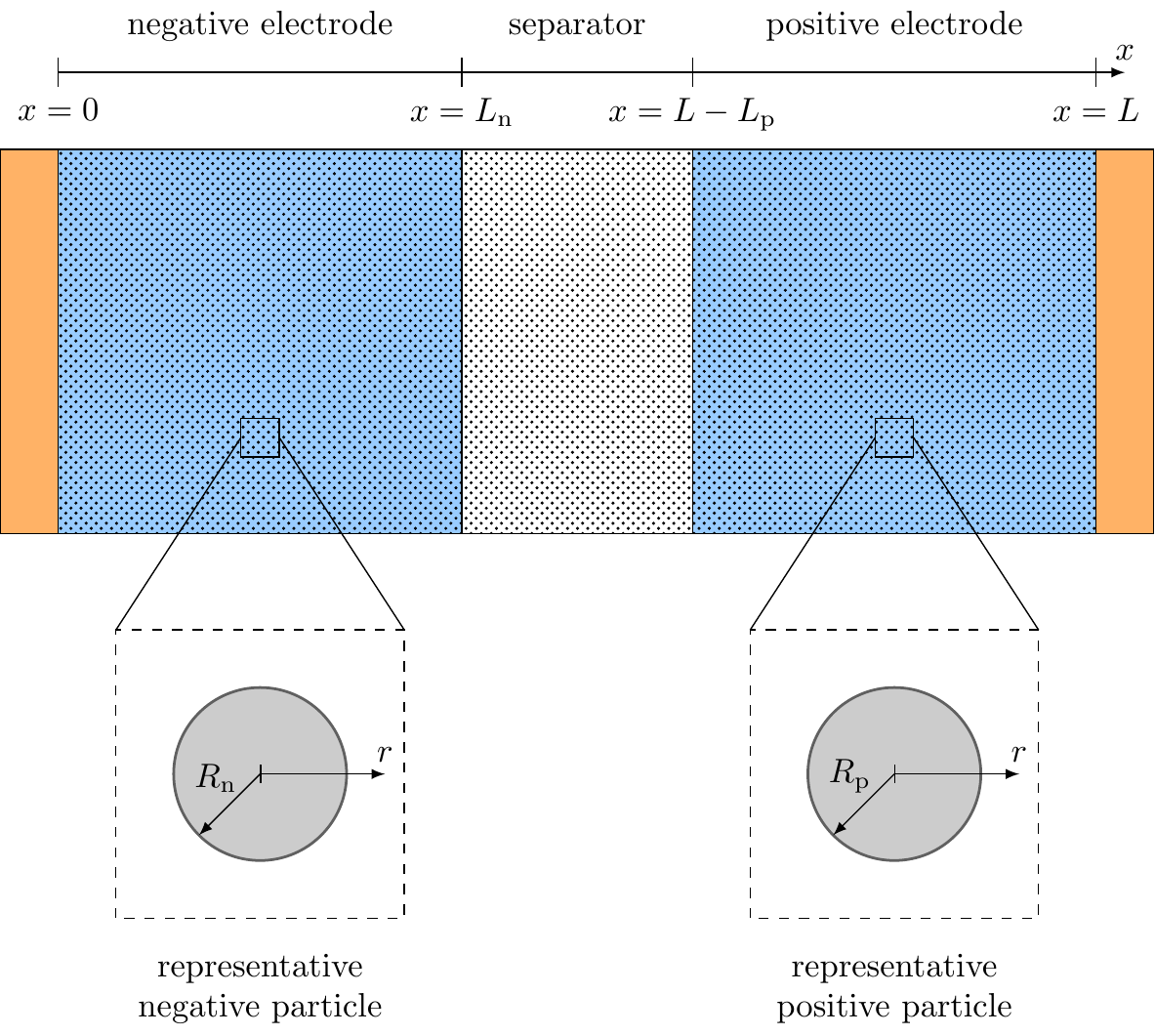}
    \caption{Geometry of the TSPMe. Note that we do not pose a geometry for the thermal model because it is a lumped model and therefore there is no space distribution for temperature. The subscripts $\mrn$ and $\mrp$ denote the parameters for the negative and positive electrodes, respectively.}
    \label{fig:geometry_sketch}
\end{figure}

The geometry of the model is shown in Figure~\ref{fig:geometry_sketch}. The two electrode particles are defined in the domain $0 \leq r \leq R_k$ (for $k \in \{\mrn,\mrp\}$), where $\mrn$ denotes the negative electrode/particle and $\mrp$ denotes the positive electrode/particle. The electrolyte domain spans across the two porous electrodes and the separator ($0 \leq x \leq L$), where the negative electrode is defined in $0 \leq x \leq L_\mrn$, the separator in $L_\mrn \leq x \leq L - L_\mrp$ and the positive electrode in $L - L_\mrp \leq x \leq L$. For the full cell, given that the heat equation is averaged over the whole cell, we do not need to define a domain.

We need to solve for the concentrations (and we are interested in certain quantities defined) both in the electrodes and the electrolyte, so we identify them with the subscripts n and p to denote the negative and positive electrodes/particles, respectively, and the subscript e to denote the electrolyte.

Then, in each particle the governing equations for intercalated lithium are 
\begin{subequations}\label{eq:SPMe_cs}
\begin{align}
\pdv{c_{k}}{t} &= \frac{1}{r^2} \pdv{}{r} \left(r^2 D_{k} \pdv{c_{k}}{r} \right), & \quad \text{ in } 0 < r < R_k,\\
\pdv{c_{k}}{r} &= 0, & \quad \text{ at } r = 0,\\
- D_{k} \pdv{c_{k}}{r} &= \frac{J_k}{a_k F}, & \quad \text{ at } r = R_k,\\
c_{k} &= c_{k,\mathrm{init}}, & \quad \text{ at } t = 0,
\end{align}
\end{subequations}
with
\begin{align}
J_\mrn &= \frac{i_\mathrm{app}}{L_\mrn}, & J_\mrp &= - \frac{i_\mathrm{app}}{L_\mrp},
\end{align}
where $c_{k}$ is the intercalated lithium concentration, $D_{k}$ is the diffusion coefficient of intercalated lithium (which may depend on $c_k$), $R_k$ is the particle radius, $J_k$ is the reaction volumetric current density, $a_k$ is the surface area density, $F$ is the Faraday constant, $c_{k,\mathrm{init}}$ is the initial concentration, $i_\mathrm{app}$ is the applied current density, and $L_k$ is the electrode thickness. Recall that the subscript $k \in \{\mathrm{n},\mathrm{p} \}$ denotes the negative and positive electrode, respectively.

The electrolyte problem is given by
\begin{subequations}\label{eq:SPMe_ce}
\begin{equation}
\varepsilon(x) \pdv{c_{\mre}}{t} = \pdv{}{x} \left( D_\mre \mathcal{B}(x) \pdv{c_{\mre}}{x} \right) + 
\begin{cases}
(1 - t^+) \frac{i_\mathrm{app}}{F L_\mrn}, & \quad \text{ if } 0 \leq x < L_\mrn,\\
0, & \quad \text{ if } L_\mrn \leq x < L - L_\mrp,\\
-(1 - t^+) \frac{i_\mathrm{app}}{F L_\mrp}, & \quad \text{ if } L - L_\mrp \leq x \leq L,
\end{cases}
\end{equation}
\begin{align}
\pdv{c_{\mre}}{x} &= 0, & \quad \text{ at } x = 0, L,\\
c_{\mre} &= c_{\mre,\mathrm{init}}, & \quad \text{ at } t = 0,
\end{align}
\end{subequations}
where $c_\mre$ is the lithium-ion concentration in the electrolyte, $\varepsilon(x)$ is the porosity (or electrolyte volume fraction), $D_\mre$ is the bulk diffusion coefficient of lithium ions (which may depend on $c_\mre$), $\mathcal{B}(x)$ is the geometry factor, $t^+$ is the transference number, $L$ is the total cell thickness, and $c_{\mre,\mathrm{init}}$ is the initial concentration. 

The spatially dependent porosity  $\varepsilon(x)$ and geometric factor $\mathcal{B}(x)$ are defined as
\begin{align}
\varepsilon(x) &= 
\begin{cases}
\varepsilon_\mrn, & \quad \text{ if } 0 \leq x < L_\mrn,\\
\varepsilon_\mrs, & \quad \text{ if } L_\mrn \leq x < L - L_\mrp,\\
\varepsilon_\mrp, & \quad \text{ if } L - L_\mrp \leq x \leq L,
\end{cases} &
\mathcal{B}(x) &= 
\begin{cases}
\mathcal{B}_\mrn, & \quad \text{ if } 0 \leq x < L_\mrn,\\
\mathcal{B}_\mrs, & \quad \text{ if } L_\mrn \leq x < L - L_\mrp,\\
\mathcal{B}_\mrp, & \quad \text{ if } L - L_\mrp \leq x \leq L,
\end{cases}
\end{align}
where the values in each subdomain $\varepsilon_k$ and $\mathcal{B}_k$ are constant. The subscripts $k \in \{\mrn, \mrs, \mrp \}$ denote the values for the negative electrode, separator and postive electrode, respectively.

The thermal problem is given by
\begin{subequations}\label{eq:TSPMe_T}
\begin{align}
\theta \dv{T}{t} &= - h a_\mathrm{cool} (T - T_\mathrm{amb}) + Q_\mrs + Q_\mre + Q_\mathrm{irr} + Q_\mathrm{rev},\\
T(0) &= T_\mathrm{amb},
\end{align}
with
\begin{align}
Q_\mrs &= \frac{i_\mathrm{app}^2}{3 L} \left( \frac{L_\mrn}{\sigma_{\mrn}} + \frac{L_\mrp}{\sigma_{\mrp}} \right),\\
Q_\mre &= - (1-t^+) \frac{2 R T}{F} \frac{i_\mathrm{app}}{L} \left( \frac{1}{L_\mrp} \int_{L-L_\mrp}^L \log \left( c_\mre \right) \dd x - \frac{1}{L_\mrn} \int_{0}^{L_\mrn} \log \left( c_\mre \right) \dd x \right) + \frac{1}{L} \int_0^L \frac{i_\mre(x,t)^2}{\sigma_\mre \left( c_\mre (x,t) \right) \mathcal{B}(x)} \dd x,\\
Q_\mathrm{irr} &= \frac{2 R T}{F} \frac{i_\mathrm{app}}{L} \left( \frac{1}{L_\mrn} \int_0^{L_\mrn} \arcsinh \left( \frac{i_\mathrm{app}}{a_\mrn L_\mrn j_\mrn}\right) \dd x + \frac{1}{L_\mrp} \int_{L-L_\mrp}^{L} \arcsinh \left( \frac{i_\mathrm{app}}{a_\mrp L_\mrp j_\mrp}\right) \dd x \right),\\
Q_\mathrm{rev} &= \frac{i_\mathrm{app}}{L} \left( \Pi_\mrn - \Pi_\mrp \right),
\end{align}
\end{subequations}
where $T$ is the average temperature of the cell (assumed to be homogeneous in space), $\theta$ is the lumped volumetric heat capacity of the cell, $h$ is the heat exchange coefficient, $a_\mathrm{cool}$ is the cooling surface area of the cell per unit of volume, $T_\mathrm{amb}$ is the ambient temperature, $Q_\mrs$ is the heat source term due to the electrode (Joule heating), $Q_\mre$ is the heat source term due to the electrolyte (both Joule heating and due to concentration gradients), $Q_\mathrm{irr}$ is the irreversible heat source term of the intercalation reaction and $Q_\mathrm{rev}$ is the reversible heat source term of the intercalation reaction. In addition, $\sigma_{k}$ is the electronic conductivity of the electrodes, $\sigma_\mre$ is the ionic conductivity of the electrolyte, $R$ is the universal gas constant, $j_k$ is the reaction surface current density, and $\Pi_k$ is the Peltier coefficient (sometimes referred to as entropic term). This coefficient is defined as $\Pi_k = T \pdv{U_k}{T}$ and describes the heat generated or sunk due to the entropy of the intercalation reaction \cite{Plett2015}.

After solving \eqref{eq:SPMe_cs}, \eqref{eq:SPMe_ce} and \eqref{eq:TSPMe_T} numerically, we can use the values of $c_k$, $c_\mre$ and $T$ to calculate any other variable of interest from explicit expressions.

For example, the potentials ($\Phi_\mrn$, $\Phi_\mrp$ and $\Phi_\mre$) and currents ($i_\mrn$, $i_\mrp$ and $i_\mre$) in the electrodes and the electrolyte can be calculated as
\begin{subequations}\label{eq:SPMe_Phi}
\begin{multline}
\Phi_{\mrn} = U_\mrn (\left. c_{\mrn} \right|_{r = R_\mrn}) - \frac{i_\mathrm{app}(2 L_\mrn - x) x}{2 L_\mrn \sigma_{\mrn}} + \frac{i_\mathrm{app} L_\mrn}{3 \sigma_{\mrn}} - \frac{1}{L_\mrn} \int_0^{L_\mrn} \int_0^x \frac{i_\mre(s,t) \dd s}{\sigma_\mre \left(c_\mre(s,t)\right) \mathcal{B}(s)} \dd x \\
+ (1 - t^+) \frac{2 R T}{F} \frac{1}{L_\mrn} \int_0^{L_\mrn} \log \left( \frac{c_\mre(x,t)}{c_\mre(0,t)}\right) \dd x + \frac{2 R T}{F} \frac{1}{L_\mrn} \int_0^{L_\mrn} \arcsinh \left( \frac{i_\mathrm{app}}{a_\mrn L_\mrn j_\mrn} \right) \dd x,\\
\end{multline}
\begin{multline}
\Phi_{\mrp} = U_\mrp (\left. c_{\mrp} \right|_{r = R_\mrp}) + \frac{i_\mathrm{app}(2 (L - L_\mrp) - x) x}{2 L_\mrp \sigma_{\mrp}} - \frac{i_\mathrm{app} \left(2 L_\mrp^2 - 6 L L_\mrp + 3 L^2 \right)}{6 L_\mrp \sigma_{\mrp}} - \frac{1}{L_\mrp} \int_{L - L_\mrp}^{L} \int_0^x \frac{i_\mre(s,t) \dd s}{\sigma_\mre \left(c_\mre(s,t)\right) \mathcal{B}(s)} \dd x \\
+ (1 - t^+) \frac{2 R T}{F} \frac{1}{L_\mrp} \int_{L-L_\mrp}^{L} \log \left( \frac{c_\mre(x,t)}{c_\mre(0,t)}\right) \dd x - \frac{2 R T}{F} \frac{1}{L_\mrp} \int_{L - L_\mrp}^{L} \arcsinh \left( \frac{i_\mathrm{app}}{a_\mrp L_\mrp j_\mrp} \right) \dd x,\\
\end{multline}
\begin{align}
\Phi_{\mre} &= (1-t^+)\frac{2 R T}{F} \log \left( \frac{c_\mre(x,t)}{c_\mre(0,t)} \right) - \int_0^x \frac{i_\mre(s,t)}{\sigma_\mre \left( c_\mre (s,t) \right) \mathcal{B}(s)} \dd s,\\
i_{\mrn} &= \frac{i_\mathrm{app}}{L_\mrn}(L_\mrn - x),\\
i_{\mrp} &= \frac{i_\mathrm{app}}{L_\mrp}(x - (L - L_\mrp)),\\
i_\mre &= \begin{cases}
\frac{i_\mathrm{app}}{L_\mrn} x, & \quad \text{ if } 0 \leq x < L_\mrn,\\
i_\mathrm{app}, & \quad \text{ if } L_\mrn \leq x < L - L_\mrp,\\
\frac{i_\mathrm{app}}{L_\mrp}(L-x), & \quad \text{ if } L - L_\mrp \leq x \leq L,
\end{cases}
\end{align}
\end{subequations}
with the exchange current densities $j_\mrn$ and $j_\mrp$ defined as
\begin{equation}
j_k = m_k \left. \sqrt{c_\mre c_{k} \left(c_{k}^{\max} - c_{k} \right)} \right|_{r = R_k}.
\end{equation}
Here, $U_k$ is the open circuit potential, $m_k$ is the intercalation reaction rate, and $c_{k}^{\max}$ is the maximum concentration in the electrode.

The output voltage is given by
\begin{subequations}\label{eq:V_TSPMe}
\begin{equation}
V_\mathrm{out} = \Phi_{\mrp}(L,t) - \Phi_{\mrn}(0,t) = U_\mathrm{eq} + \eta_\mathrm{r} + \eta_\mre + \Delta \Phi_\mre + \Delta \Phi_\mrs,
\end{equation}
where
\begin{align}
U_\mathrm{eq} &= U_\mrp (\left.c_{\mrp}\right|_{r = R_\mrp}) - U_\mrn (\left. c_{\mrn}\right|_{r = R_\mrn}),\\
\eta_\mathrm{r} &= - \frac{2 R T}{F} \left(\frac{1}{L_\mrp}\int_{L-L_\mrp}^{L} \arcsinh \left( \frac{i_\mathrm{app}}{a_\mrp L_\mrp j_\mrp} \right) \dd x + \frac{1}{L_\mrn} \int_{0}^{L_\mrn} \arcsinh \left( \frac{i_\mathrm{app}}{a_\mrn L_\mrn j_\mrn} \right) \dd x \right),\\
\eta_\mre &= (1-t^+) \frac{2 R T}{F} \left( \frac{1}{L_\mrp} \int_{L-L_\mrp}^L \log \left( c_\mre \right) \dd x - \frac{1}{L_\mrn} \int_{0}^{L_\mrn} \log \left( c_\mre \right) \dd x \right),\\
\Delta \Phi_\mre &= - \left(\frac{1}{L_\mrp} \int_{L - L_\mrp}^{L} \int_0^x \frac{i_\mre(s,t) \dd s}{\sigma_\mre \left(c_\mre(s,t)\right) \mathcal{B}(s)} \dd x - \frac{1}{L_\mrn} \int_0^{L_\mrn} \int_0^x \frac{i_\mre(s,t) \dd s}{\sigma_\mre \left(c_\mre(s,t)\right) \mathcal{B}(s)} \dd x \right),\\
\Delta \Phi_\mrs &= - \frac{i_\mathrm{app}}{3} \left( \frac{L_\mrn}{\sigma_{\mrn}} + \frac{L_\mrp}{\sigma_{\mrp}} \right).
\end{align}
\end{subequations}
We can interpret these terms in the following way: $U_\mathrm{eq}$ is the equilibrium potential to which the terminal voltage converges when no current is applied, while all the other terms are deviations from equilibrium due to different effects. $\eta_\mathrm{r}$ is due to the reaction overpotentials, $\eta_\mre$ is due to the concentration gradients in the electrolyte, $\Delta \Phi_\mre$ is due to Ohmic losses in the electrolyte, and $\Delta \Phi_\mrs$ is due to Ohmic losses in the (solid) electrode.

The TSPMe is valid in a particular regime both in terms of electrochemical and thermal submodels, and the physical intuition for them is discussed in Section \ref{sec:SPM_discussion}. The electrochemical submodel is valid when the deviations from the equilibrium potential are slow, which is equivalent to moderate to low C-rates. In terms of the dimensionless parameters, this corresponds to the case when
\begin{align}
    \lambda &= \frac{F \Phi_0}{R T_\mathrm{amb}} \gg 1, & \Sigma_k &= \frac{R T_\mathrm{amb}}{F L i_0} \sigma_k \gtrsim 1, & \Sigma_\mre &= \frac{R T_\mathrm{amb}}{F L i_0} \sigma_{\mre,\typ} \gtrsim 1,
\end{align}
where $i_0$ and $\Phi_0$ are the typical current and electrode potential, respectively, and $\sigma_{\mre,\typ}$ is the typical value of $\sigma_\mre$. The assumption $\lambda \gg 1$ (i.e. $\lambda$ significantly larger than one) means that $\lambda$ must be larger than one for the assumption to hold and, the larger $\lambda$ is, the better the reduced model works. On the other hand, the assumption of $\Sigma_k \gtrsim 1$ (and similarly for $\Sigma_\mre$) means that $\Sigma_k$ must be of $\order{1}$ or larger (i.e. not significantly smaller than one). Further details on these dimensionless parameters, as well as the details of the asymptotic reduction, can be found in \ref{sec:derivation_TSPMe}.

The thermal submodel is valid when the heat exchange with environment is much slower than the heat transfer inside of the battery. In terms of dimensionless parameters, this corresponds to the case when
\begin{align}
    \Bi &= \frac{h L_\mathrm{batt}}{\kappa} \ll 1, & \mathcal{K} &= \frac{\kappa t_0}{L^2_\mathrm{batt} \theta} \gg 1,
\end{align}
where $L_\mathrm{batt}$ is the typical length scale of the battery, $\kappa$ is the lumped thermal conductivity of the battery and $t_0$ is the typical discharge time.

To the best of our knowledge, the TSPMe presented here is the first instance of a formal derivation (i.e. using asymptotic techniques) of a TSPM-type model for multi-layer batteries and including finite diffusion effects in the particles. There are other instances of asymptotically derived thermal-electrochemical models in the literature, but they focus on other aspects. The work in \cite{Marquis2020,Timms2020} focuses on single layer pouch cells. On the other hand, the analysis in \cite{Hennessy2019} assumes infinitely fast diffusion, which does capture the relaxation of voltage that we observe in the experimental data when the battery is switched off (see Section \ref{sec:results}). 

In terms of the electrochemical model, our analysis is based on the same assumptions as the work by Richardson et al. \cite{Richardson2020}, but in our case we focus on the homogeneous electrodes case. Therefore, we are able to write the electrolyte potential (\ref{eq:SPMe_Phi}c) as an explicit expression rather than having to solve an elliptic PDE for it. This allows us to write simpler expressions for all the potentials, while retaining all the features in \cite{Richardson2020} for homogeneous electrodes. On the other hand, compared to the model by Marquis et al. \cite{Marquis2019}, our approach directly captures all the nonlinearities in the voltage expression, so there is no need to introduce them \emph{a posteriori} (see \cite{Marquis2020} for details).

\subsection{Further simplifications}\label{sec:further_simplifications}
Notice that the TSPMe is already an SPM-type model, as defined in Section \ref{sec:SPM_discussion}, which can be implemented at a very low computational cost. Therefore, from the point of view of computational efficiency, there is no need to further simplify the model. However, in the literature we find further simplifications to SPM-type models. These further simplified models can also be helpful to understand the physical behaviour of batteries and for applications where the available computational power is really low (such as in battery management systems). In this section we present some further simplifications that can be done by taking additional assumptions to show the links between the TSPMe model and other models found in the literature \cite{Marquis2019,Richardson2020,Marquis2020,Richardson2012,Moyles2019asymptotic,Moyles2018experimental,Hennessy2019}. The full details of the derivation can be found in \ref{sec:further_simplifications_app}.

Here we present the dimensional form of three main different simplifications: quasi-steady-state electrolyte concentration, constant electrolyte conductivity and fast lithium diffusion. For the first two we consider as well a particular case of each that allows even further reduction of the model (small variation of electrolyte concentration and large electrolyte conductivity). Notice that the three simplifications are independent from each other, so we can choose to use a subset of them or all together.

\subsubsection{Quasi-steady state electrolyte concentration}
One simplification for the electrolyte concentration is to take the quasi-steady-state problem, which is a valid assumption when the current varies over a larger time scale than the electrolyte diffusion and electrolyte ion generation time scales. Mathematically, this corresponds to the case when
\begin{align}
    \mathcal{C}_\mre &= \frac{L^2}{D_{\mre,\typ} t_0} \ll 1, & \gamma_\mre &= \frac{c_{\mre,\mathrm{init}}}{c_\mrn^{\max}} \ll 1,
\end{align}
where $D_{\mre,\typ}$ is the typical value of $D_\mre$ and $c_\mrn^{\max}$ is the maximum concentration in the negative electrode particles.

The physical meaning of this limit is that electrolyte transient effects are negligible because they happen at a much shorter time scale than variations in the applied current. This is a reasonable assumption for constant-current discharge (or charge) experiments.

With these assumptions, there is no time dependence in the electrolyte concentration equations so their integration is a one-off step. Additionally, if the ion transport properties in the electrolyte ($D_\mre$ and $t^+$) are assumed to be constant, we can find the following analytical expression for the concentration
\begin{subequations}\label{eq:ce_ss}
\begin{equation}
    c_\mre = c_{\mre,\mathrm{init}} + \frac{i_\mathrm{app} (1 - t^+)}{6 F D_\mre v_\mathrm{pore}} \Delta c (x)
\end{equation}
with
\begin{equation}
    \Delta c (x) = \begin{cases}
        \left(\frac{2 \varepsilon_\mrp L_\mrp^2}{\mathcal{B}_\mrp} + \frac{3 L_\mrs (2 \varepsilon_\mrp L_\mrp + \varepsilon_\mrs L_\mrs)}{\mathcal{B}_\mrs} - \frac{3 \frac{v_\mathrm{pore}}{L_\mrn} (L_\mrn^2 - x^2) - 2 \varepsilon_\mrn L_\mrn^2}{\mathcal{B}_\mrn} \right), & \text{ if } 0 < x < L_\mrn,\\
        \left( - \frac{2 \varepsilon_\mrn L_\mrn^2}{\mathcal{B}_\mrn} + \frac{2 \varepsilon_\mrp L_\mrp^2}{\mathcal{B}_\mrp} + \frac{6 \frac{v_\mathrm{pore}}{L} (L - L_\mrp - x) - 6 \varepsilon_n L_\mrn L_\mrs - 3 \varepsilon_\mrs L_\mrs^2}{\mathcal{B}_\mrs}\right), & \text{ if } L_\mrn < x < L - L_\mrp,\\
        \left( - \frac{2 \varepsilon_\mrn L_\mrn^2}{\mathcal{B}_\mrn} - \frac{3 L_\mrs (2 \varepsilon_\mrn L_\mrn + \varepsilon_\mrs L_\mrs)}{\mathcal{B}_\mrs} + \frac{3 \frac{v_\mathrm{pore}}{L_\mrp} \left( (L^2 - L_\mrp^2) - (2L -x) x \right) + 2 \varepsilon_\mrp L_\mrp^2}{\mathcal{B}_\mrp}\right), & \text{ if } L - L_\mrp < x < L,
    \end{cases}
\end{equation}
\end{subequations}
where the volume of pores per unit of electrode plate area is given by
\begin{equation}
v_\mathrm{pore} = \int_0^L \varepsilon(x) \dd x  = \varepsilon_\mrn L_\mrn + \varepsilon_\mrs L_\mrs + \varepsilon_\mrp L_\mrp,
\end{equation}
and $L_\mrs = L - L_\mrn - L_\mrp$ is the thickness of the separator.

An additional step to the previous simplification occurs when the variation in electrolyte concentration is small. This corresponds to the limit
\begin{align}
    \mathcal{C}_\mre &= \frac{L^2}{D_{\mre,\typ} t_0} \ll 1, & \gamma_\mre &= \frac{c_{\mre,\mathrm{init}}}{c_\mrn^{\max}} \sim 1,
\end{align}
and it yields the result $c_\mre = c_{\mre,\mathrm{init}}$. Therefore, we obtain a Single Particle Model as we do not need to solve for $c_{\mre}$. Notice that this limit is the same to the one studied in \cite{Marquis2019} in terms of electrolyte concentration. The only differences between that model and the one presented here are in the voltage expressions. In their case they assumed large conductivity in the electrodes, while we assume small overpotentials (i.e. $\lambda \gg 1$), but the results are analogous.  

This assumption also helps simplify the expression for the electrolyte potential because we find that $\sigma_\mre (c_\mre)$ is a constant, which is the case discussed in the next section.

\subsubsection{Constant electrolyte conductivity}\label{sec:constant_sigmae}
One assumption to simplify the electrolyte potential is to take that the electrolyte conductivity is a constant. Then, the integrals involving $\sigma_\mre$ in (\ref{eq:TSPMe_T}c), (\ref{eq:SPMe_Phi}) and (\ref{eq:V_TSPMe}) can be computed analytically. This means a slight reduction in the computational cost, which is negligible for a laptop but can be important for simpler devices such as the processors in the battery management systems of the electric vehicles. 

Hence, the potentials in both electrode and electrolyte can be calculated using
\begin{subequations}
\begin{multline}
\Phi_{\mrn} = U_\mrn (\left.c_{\mrn}\right|_{r = R_\mrn}) - \frac{i_\mathrm{app}(2 L_\mrn - x) x}{2 L_\mrn \sigma_{\mrn}} + \frac{i_\mathrm{app} L_\mrn}{3 \sigma_{\mrn}} - \frac{i_\mathrm{app}}{6 \sigma_\mre} \frac{L_\mrn}{\mathcal{B}_\mrn} \\
+ (1 - t^+) \frac{2 R T}{F} \frac{1}{L_\mrn} \int_0^{L_\mrn} \log \left( \frac{c_\mre(x,t)}{c_\mre(0,t)}\right) \dd x + \frac{2 R T}{F} \frac{1}{L_\mrn} \int_0^{L_\mrn} \arcsinh \left( \frac{i_\mathrm{app}}{a_\mrn L_\mrn j_\mrn} \right) \dd x,
\end{multline}
\begin{multline}
\Phi_{\mrp} = U_\mrp (\left.c_{\mrp}\right|_{r = R_\mrp}) + \frac{i_\mathrm{app}(2 (L - L_\mrp) - x) x}{2 L_\mrp \sigma_{\mrp}} - \frac{i_\mathrm{app} \left(2 L_\mrp^2 - 6 L L_\mrp + 3 L^2 \right)}{6 L_\mrp \sigma_{\mrp}} - \frac{i_\mathrm{app}}{6 \sigma_\mre} \left(\frac{3 L_\mrn}{\mathcal{B}_\mrn} + \frac{6 L_\mrs}{\mathcal{B}_\mrs} + \frac{2 L_\mrp}{\mathcal{B}_\mrp} \right)\\
+ (1 - t^+) \frac{2 R T}{F} \frac{1}{L_\mrp} \int_{L-L_\mrp}^{L} \log \left( \frac{c_\mre(x,t)}{c_\mre(0,t)}\right) \dd x - \frac{2 R T}{F} \frac{1}{L_\mrp} \int_{L - L_\mrp}^{L} \arcsinh \left( \frac{i_\mathrm{app}}{a_\mrp L_\mrp j_\mrp} \right) \dd x,
\end{multline}
\begin{align}
\Phi_{\mre} &= (1-t^+)\frac{2 R T}{F} \log \left( \frac{c_\mre(x,t)}{c_\mre(0,t)} \right) + \frac{i_\mathrm{app}}{2 \sigma_\mre}
\begin{cases}
-\frac{x^2}{\mathcal{B}_\mrn L_\mrn}, & \quad \text{ if } 0 \leq x < L_\mrn,\\
-\frac{2 (x - L_\mrn)}{\mathcal{B}_\mrs} - \frac{L_\mrn}{\mathcal{B}_\mrn}, & \quad \text{ if } L_\mrn \leq x < L - L_\mrp,\\
\frac{(L-x)^2}{\mathcal{B}_\mrp L_\mrp} - \left(\frac{L_\mrn}{\mathcal{B}_\mrn} + \frac{2 L_\mrs}{\mathcal{B}_\mrs} + \frac{L_\mrp}{\mathcal{B}_\mrp} \right), & \quad \text{ if } L - L_\mrp \leq x \leq L.
\end{cases}
\end{align}
\end{subequations}

Then, the Ohmic losses in the electrolyte can be written as
\begin{equation}
\Delta \Phi_\mre = - \frac{i_\mathrm{app}}{3 \sigma_\mre} \left( \frac{L_\mrn}{\mathcal{B}_\mrn} + \frac{3 L_\mrs}{\mathcal{B}_\mrs} + \frac{L_\mrp}{\mathcal{B}_\mrp} \right),
\end{equation}
and the electrolyte heat source term can be written as
\begin{equation}
Q_\mre = - (1-t^+) \frac{2 R T}{F} \frac{i_\mathrm{app}}{L} \left( \frac{1}{L_\mrp} \int_{L-L_\mrp}^L \log \left( c_\mre \right) \dd x - \frac{1}{L_\mrn} \int_{0}^{L_\mrn} \log \left( c_\mre \right) \dd x \right) + \frac{i_\mathrm{app}^2}{3 L \sigma_\mre} \left( \frac{L_\mrn}{\mathcal{B}_\mrn} + \frac{3 L_\mrs}{\mathcal{B}_\mrs} + \frac{L_\mrp}{\mathcal{B}_\mrp} \right).
\end{equation}

Note that, as mentioned in the previous section, these expressions for the potentials are very similar to those obtained in the limit where the variation of electrolyte concentration is small. Therefore, we can observe that the Ohmic losses in the electrolyte have the same form as those in \cite{Marquis2019}.

If the electrolyte conductivity is assumed to be high, then an additional simplification can be done by eliminating all the terms involving $\sigma_\mre^{-1}$. Mathematically, this corresponds to the limit
\begin{equation}
    \Sigma_\mre = \frac{R T_\mathrm{amb}}{F L i_0} \sigma_{\mre,\typ} \gg 1.
\end{equation}

\subsubsection{Fast electrode diffusion}\label{sec:fast_diffusion}
Finally, the last simplification that can be taken is fast electrode diffusion, which corresponds to the limit
\begin{equation}
    \mathcal{C}_k = \frac{R^2_k}{D_{k,\typ} t_0} \ll 1.
\end{equation}
In this case, the particle problem \eqref{eq:SPMe_cs} reduces to an ODE of the form
\begin{subequations}
\begin{align}
    \dv{c_k}{t} &= -\frac{J_k}{a_k R_k F}, \\
    c_k(0) &= c_{k,\mathrm{init}},
\end{align}
\end{subequations}
and we find several instances of this model in the literature, such as \cite{Richardson2012,Moyles2019asymptotic,Moyles2018experimental,Hennessy2019}. Note that this fast electrode diffusion model does not capture the relaxation effects that we observe when the current is switched off. These effects, mostly noticeable in the terminal voltage, are caused by the relaxation of the particle concentration to its steady state and therefore require a finite diffusion in the particles.

\section{Results}\label{sec:results}
After presenting the TSPMe model, we now validate it to show its relevance and applicability. To show the accuracy and performance of the reduced model, we start by comparing the TSPMe with the thermal DFN (TDFN). Then, to show the applicability of the TSPMe in real applications we compare it against experimental data on the LG M50 cell. This is a 21700 cylindrical cell, with a nominal capacity of 5 Ah, and with an NMC811 positive electrode and a graphite and SiO$_x$ negative electrode (see \cite{Chen2020} for further details). In both cases, we perform comparisons at different C-rates (C/2, 1C and 2C) and ambient temperatures (0~$\degC$, 10~$\degC$ and 25~$\degC$). The code to reproduce the simulations is publicly available in an online repository (\url{https://github.com/brosaplanella/TEC-reduced-model}).

\subsection{Comparison between TSPMe and DFN}
We first look at the comparison between the reduced model (the TSPMe) and a thermal DFN model (using a lumped thermal model, see \cite{Marquis2020}) under different discharge conditions to assess how well does the TSPMe approximate the TDFN. Here we use a lumped thermal model coupled to the DFN for simplicity, as resolving the battery geometry would require a way heavier computational approach that is out of the scope of this work. This increase of computational cost is because, as detailed in \cite{Tranter2020}, we would have to define a model over three scales (battery, cell and particle, see Figure \ref{fig:multiscale}) with the corresponding battery geometry, which would also be challenging from the implementation point of view. The price we pay to use a lumped model for the TDFN is that we do not capture the thermal gradients in the battery. Validating the TSPMe model against a spatially resolved TDFN (like \cite{Tranter2020}) is an area for future work on this topic.

The models have been implemented in PyBaMM (Python Battery Mathematical Modelling) package, a Python-based open-source package to simulate battery models \cite{Sulzer2020}. The code and data are available online (see ``Data and code availability''). Both models are solved using a finite volumes method for the spatial discretisation using the same number of points in both methods for a given domain. For the results here, we used 30 points in the electrode particles and 20 points for each electrode and separator. With this number of points in the discretisation the model yields accurate results, and we do not observe a significant improvement in the solution when increasing the number of points in the mesh. The code to compare the different mesh sizes can be found in the online repository.

After spatial discretisation, we obtain a system of 122 ODEs for the TSPMe model and a system of 1262 ODEs and 100 algebraic equations for the TDFN model, which is an index one DAE \cite{Marquis2019}. Observe that, for the same mesh size, the TDFN model requires solving a system over 10 times larger than the TSPMe, and much more complex due to the algebraic constraints. In order to solve the models we used the SciPy ODE solver \cite{Virtanen2020} for the TSPMe, while for the TDFN we needed a DAE solver, so we used the CasADI solver \cite{Andersson2019}.

\begin{table}
\centering
\begin{tabular}{| c c l c c c |}
\hline
\textbf{Symbol} & \textbf{Units} & \textbf{Description} & \textbf{Pos.} & \textbf{Sep.} & \textbf{Neg.} \\ \hline
$L_{k}$ & m & Thickness & $75.6\E{-6}$ & $12\E{-6}$ & $85.2\E{-6}$ \\
$R_{k}$ & m & Radius of electrode particles & $5.22\E{-6}$ & - & $5.86\E{-6}$ \\
$a_{k}$ & m$^{-1}$ & Particle surface area density & $3.82\E{5}$ & - & $3.84\E{5}$ \\
$D_{k}$ & $\mathrm{m}^2 \; \mathrm{s}^{-1}$ & Lithium diffusivity in particles & $4\E{-15}$ & - & $3.3\E{-14}$ \\
$\sigma_{k}$ & $\mathrm{S} \; \mathrm{m}^{-1}$ & Electrode conductivity & 0.18 & - & 215 \\
$c_{k,\mathrm{init}}$ & $\mathrm{mol} \; \mathrm{m}^{-3}$ & Initial particle concentration & 17038 & - & 29866 \\
$c_{k}^{\max}$ & $\mathrm{mol} \; \mathrm{m}^{-3}$ & Max. particle concentration & 63104 & - & 33133 \\
$U_k$ & V & Open circuit potential & Fig. \ref{fig:OCVs} & - & Fig. \ref{fig:OCVs} \\
$m_k$ & $\mathrm{A} \; \mathrm{m}^{-2} \left(\mathrm{mol} \; \mathrm{m}^{-3}\right)^{-1.5}$ & Reaction rate & $3.42\E{-6}$ & - & $6.48\E{-7}$ \\ \hline
$\varepsilon_{k}$ & - & Electrolyte volume fraction & 0.335 & 0.47 & 0.25 \\
$D_{\mre}$ & $\mathrm{m}^2 \; \mathrm{s}^{-1}$ & Electrolyte diffusivity & \multicolumn{3}{c |}{see \cite{Chen2020}} \\
$\sigma_{\mre}$ & $\mathrm{S} \; \mathrm{m}^{-1}$ & Electrolyte conductivity & \multicolumn{3}{c |}{see \cite{Chen2020}} \\
$t^+$ & - & Transfer number & \multicolumn{3}{c |}{0.2594} \\
$c_{\mre,\mathrm{init}}$ & $\mathrm{mol} \; \mathrm{m}^{-3}$ & Initial electrolyte concentration & \multicolumn{3}{c |}{1000} \\ \hline
$i_{\mathrm{app}}$ & $\mathrm{A} \; \mathrm{m}^{-2}$ & Applied current density & \multicolumn{3}{c |}{$48.69 C$}\\
$F$ & $\mathrm{C} \; \mathrm{mol}^{-1}$ & Faraday constant & \multicolumn{3}{c |}{96485}\\
$R$ & $\mathrm{J} \; \mathrm{K}^{-1} \; \mathrm{mol}^{-1}$ & Gas constant & \multicolumn{3}{c |}{8.314}\\
$a_\mathrm{cool}$ & $\mathrm{m}^{-1}$ & Cooling surface area density & \multicolumn{3}{c |}{219.42} \\
$L_\mathrm{batt}$ & m & Length scale of the battery* & \multicolumn{3}{c |}{$1\E{-2}$}\\ 
$T_\mathrm{amb}$ & K & Ambient temperature & \multicolumn{3}{c |}{298}\\ \hline
$\theta$ & $\mathrm{J} \; \mathrm{K}^{-1} \; \mathrm{m}^{-3}$ & Volumetric heat capacity & \multicolumn{3}{c |}{$2.85\E{6}$} \\
$\kappa$ & $\mathrm{W} \; \mathrm{m}^{-1} \; \mathrm{K}^{-1}$ & Thermal conductivity* & \multicolumn{3}{c |}{1.05} \\
$h$ & $\mathrm{W} \; \mathrm{m}^{-2} \; \mathrm{K}^{-1}$ & Heat exchange coefficient & \multicolumn{3}{c |}{20} \\ \hline
\end{tabular}
\caption{Dimensional parameters for the LG M50 cell. The electrochemical parameters are taken from \cite{Chen2020} and the thermal parameters are taken from \cite{Taheri2013}. The parameters marked with an asterisk indicate that they are not directly used in the TSPMe but have been used to evaluate the dimensionless grouping for its derivation (see \ref{sec:derivation_TSPMe}). The $C$ in the definition of $i_\mathrm{app}$ corresponds to the C-rate of the experiment. The open-circuit potentials of the electrodes are taken from \cite{Chen2020} as well, and they are plotted in Figure \ref{fig:OCVs}.}
\label{tab:parameter_values_LGM50}
\end{table}

For the simulations we used the parameter values shown in Table~\ref{tab:parameter_values_LGM50}. The electrochemical parameters are for the LG M50 cell  and have been taken from \cite{Chen2020}, while the thermal parameters ($\theta$, $\kappa$ and $h$) have been taken from \cite{Taheri2013}. For the thermal parameters, given that they are lumped parameters for the whole battery, we took from \cite{Taheri2013} the values for each part and calculated the lumped value using the thicknesses from \cite{Chen2020}. For the heat exchange coefficient, we took the intermediate value from the range discussed in \cite{Taheri2013}. Note that two of the parameters in Table~\ref{tab:parameter_values_LGM50} ($L_\mathrm{batt}$ and $\kappa$) do not directly appear in the TSPMe model, but their values have been used to calculate the dimensionless parameters in Table~\ref{tab:ND_parameter_values_LGM50}, which are needed in the asymptotic analysis presented in \ref{sec:derivation_TSPMe}. These parameters are identified with an asterisk in Table~\ref{tab:parameter_values_LGM50}. To capture the microstructure effects on the macroscopic transport, we use the Bruggeman correlation, so we set $\mathcal{B}_k = \varepsilon_k^{1.5}$. We also do not include the reversible heat generation effects $Q_\mathrm{rev}$ because the Peltier coefficient $\Pi_k = T \pdv{U_k}{T}$ needs to be consistent with the open-circuit potential $U_k$ and this would require a careful and exhaustive experimental analysis that is out of the scope of this work. This is an aspect that will be addressed in future work.

\begin{figure}
    \centering
    \includegraphics[scale=1]{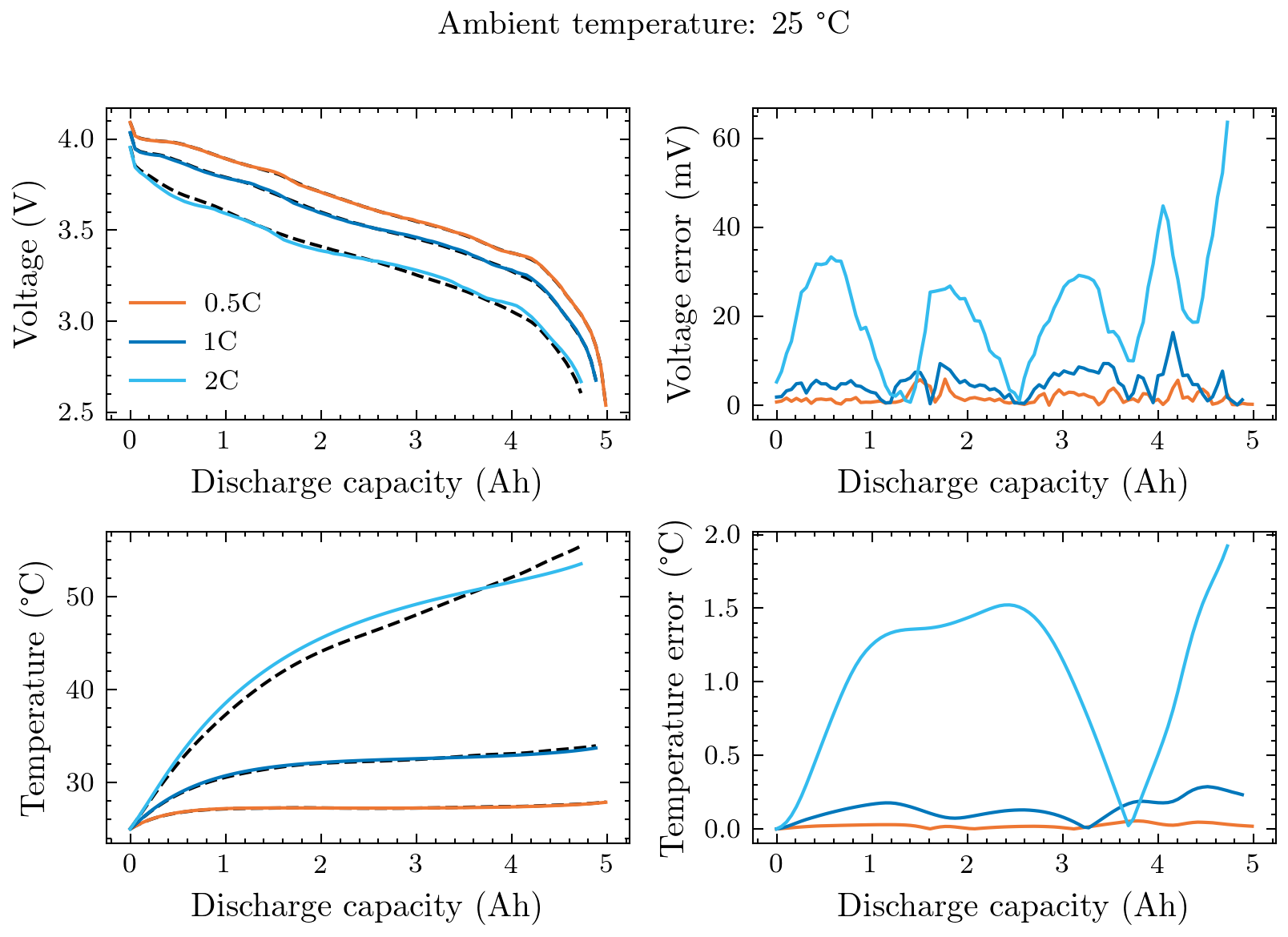}
    \caption{Comparison between the TDFN and TSPMe solutions at 25 $\degC$ and different C-rates. The plots on the left column compare the terminal voltage and cell temperature, where the solid colour lines represent the TSPMe and the black dashed lines represent the TDFN. The plots on the right column show the absolute error between the two models for voltage and temperature, respectively.}
    \label{fig:comp_models_25degC}
\end{figure}

\begin{figure}
    \centering
    \includegraphics[scale=1]{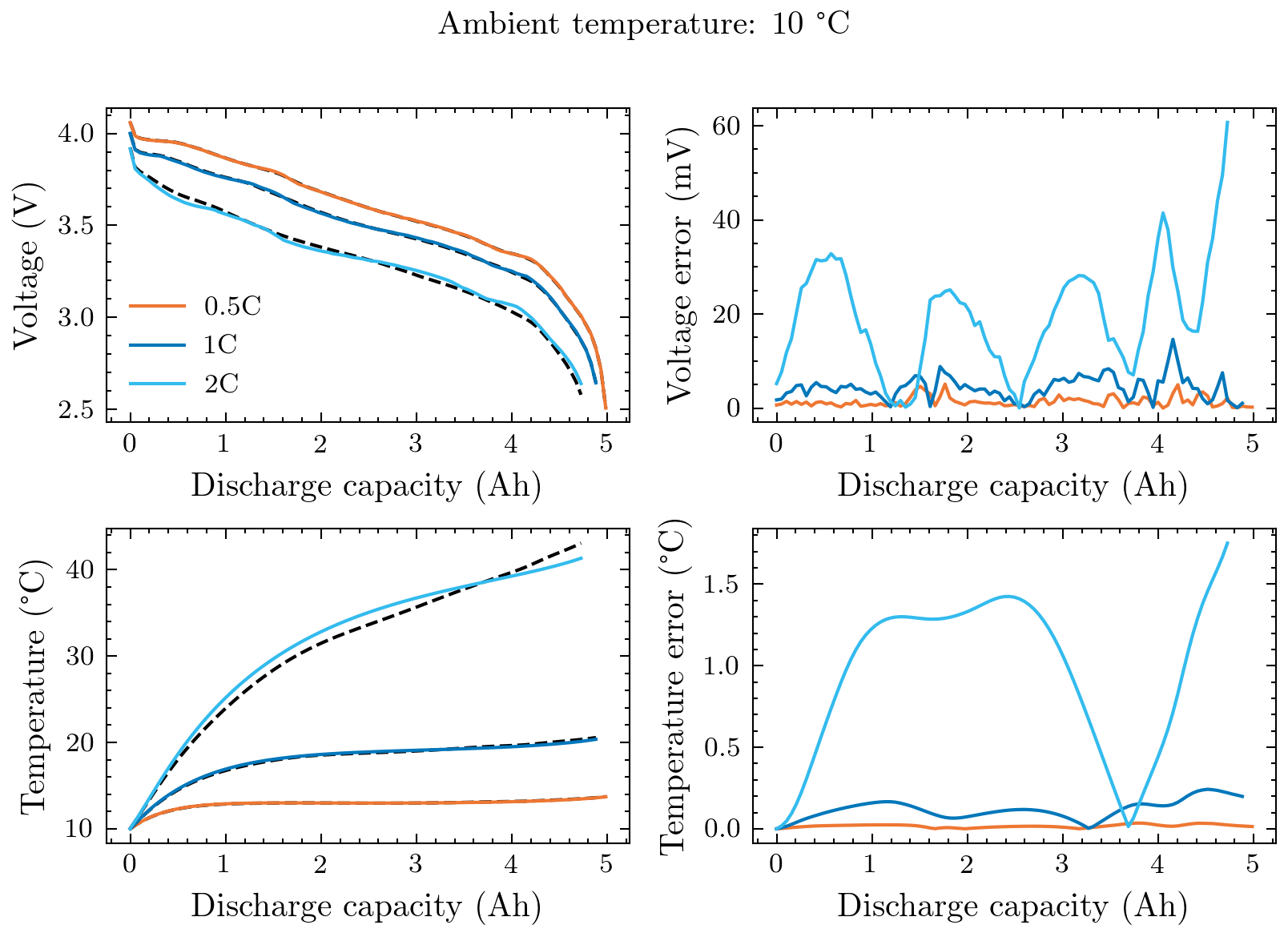}
    \caption{Comparison between the TDFN and TSPMe solutions at 10 $\degC$ and different C-rates. The plots on the left column compare the terminal voltage and cell temperature, where the solid colour lines represent the TSPMe and the black dashed lines represent the TDFN. The plots on the right column show the absolute error between the two models for voltage and temperature, respectively.}
    \label{fig:comp_models_10degC}
\end{figure}

\begin{figure}
    \centering
    \includegraphics[scale=1]{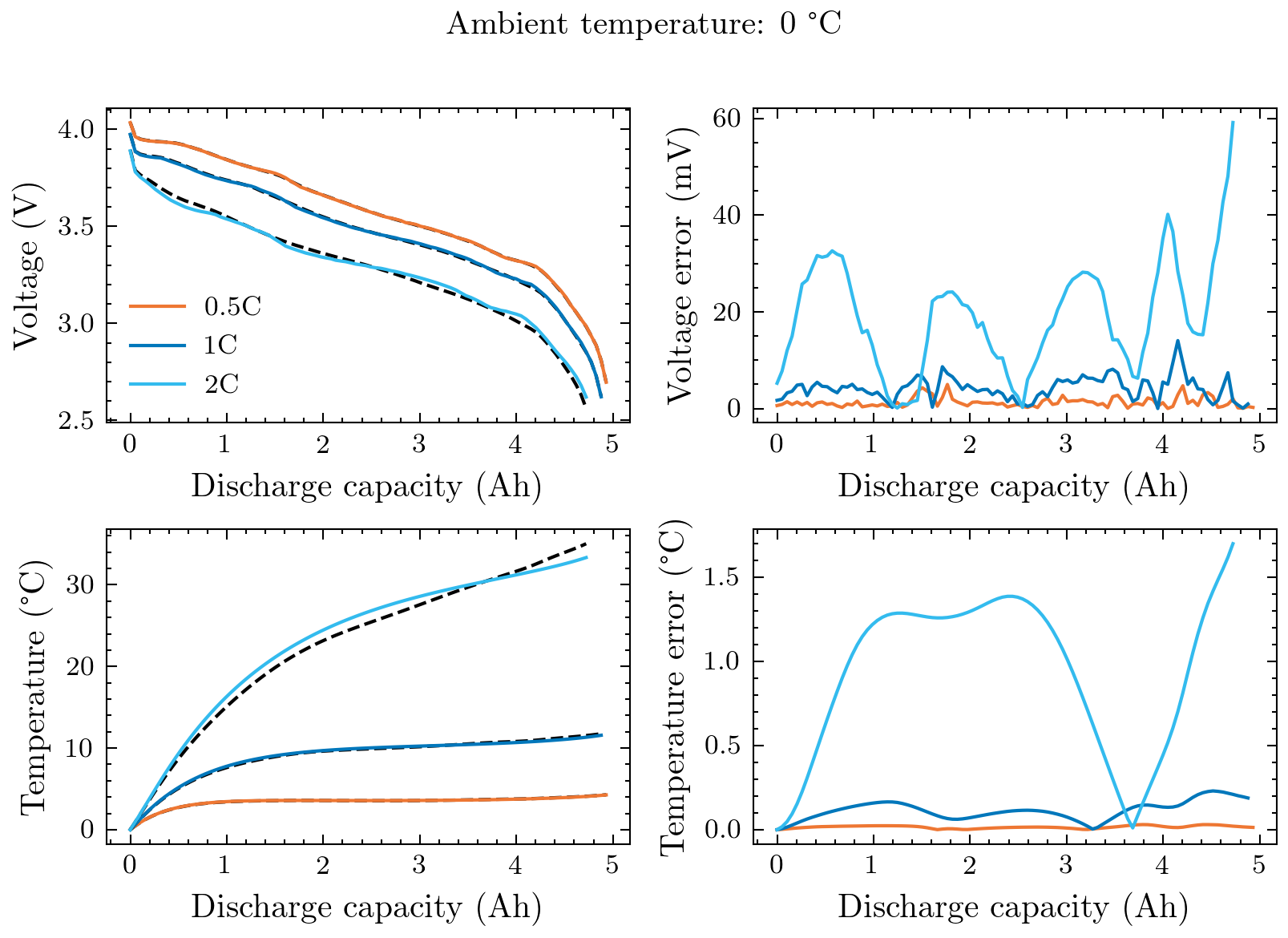}
    \caption{Comparison between the TDFN and TSPMe solutions at 0 $\degC$ and different C-rates. The plots on the left column compare the terminal voltage and cell temperature, where the solid colour lines represent the TSPMe and the black dashed lines represent the TDFN. The plots on the right column show the absolute error between the two models for voltage and temperature, respectively.}
    \label{fig:comp_models_0degC}
\end{figure}

The comparison between the TDFN model and TSPMe is shown in Figures~\ref{fig:comp_models_25degC}-\ref{fig:comp_models_0degC}, with the error data in Table~\ref{tab:error_models} and the computational times in Table~\ref{tab:comp_time}. We observe that the TSPMe does a very good job representing the voltage and the cell temperature compared to the TDFN model, and that the discrepancy increases with the C-rate. This could be expected as, from the asymptotic analysis in \ref{sec:derivation_TSPMe}, we know that some of the assumptions on the parameter sizes break down if the applied current is too large. The data in Table~\ref{tab:error_models} shows a slight trend that the agreement between models improves as the temperature decreases, which could be caused because lower temperatures yield lower values of the thermal potential, which is one of the assumptions in the asymptotic analysis. 

Meanwhile, the computational time to solve each model shows that the TSPMe is between approximately 10 times (2C) and 40 times (C/2) faster than the TDFN model. We also notice that, for the TSPMe, the computational time does not vary much with the C-rate, but for the TDFN model the computational time is inversely proportional to the C-rate. This is due to the way PyBaMM handles solver events (e.g. stopping the solver when the critical discharge voltage is reached). The parameters of the solver could be changed to reduce the solving time of the TDFN, but they would need to be adjusted in a case by case basis given that the solver is very sensitive to the parameter values and the experimental conditions. On the other hand, there is no significant effect of the experiment temperature on the computational time. The computational times have been calculated on a laptop with an Intel Core i7-7660U (2.50Ghz) processor and 16 GB RAM. To eliminate the artefacts caused by other background processes, each model has been solved 20 times and the values shown in Table~\ref{tab:comp_time} are the mean and standard deviation of the different runs for each C-rate and temperature. 

This huge difference in the computational speed is because the PDEs arising in the TSPMe are discretised into a system of ODEs which is well-behaved, while the discretised TDFN model gives a DAE system which is ill-posed and, therefore, solving this system requires very specific numerical schemes that are computationally expensive \cite{Gerdts2015}. Specialised techniques to speed up solving both systems of equations are available (like the reduction methods of residue grouping \cite{Smith2008} or balanced truncation \cite{Jun2015} for systems of ODEs). However, given that the TSPMe is intrinsically simpler than the TDFN, under equivalent conditions the former should always be faster. From this comparison, we conclude that the TSPMe provides a very good approximation to the thermal DFN model at a much lower computational cost. In the next section we assess if this model can be used to obtain meaningful predictions of experimental data.

\begin{table}
    \begin{subtable}{.5\linewidth}
      \centering
        \begin{tabular}{r | c | c | c}
            & 0.5C & 1C & 2C \\ \hline
            25 $\degC$ & 2.10 (5.87) & 5.59 (16.35) & 23.95 (63.61) \\
            10 $\degC$ & 1.72 (5.10) & 4.97 (14.62) & 22.58 (60.71) \\
            0 $\degC$ & 1.64 (4.98) & 4.82 (14.05) & 22.10 (59.15) \\
        \end{tabular}
        \caption{Voltage (mV)}
    \end{subtable}%
    \begin{subtable}{.5\linewidth}
      \centering
        \begin{tabular}{r | c | c | c}
            & 0.5C & 1C & 2C \\ \hline
            25 $\degC$ & 0.03 (0.05) & 0.15 (0.29) & 1.14 (1.92) \\
            10 $\degC$ & 0.02 (0.04) & 0.13 (0.24) & 1.07 (1.75) \\
            0 $\degC$ & 0.02 (0.03) & 0.13 (0.23) & 1.04 (1.70) \\
        \end{tabular}
        \caption{Temperature ($\degC$)}
    \end{subtable}
    \caption{Error between TSPMe and DFN models for different temperatures and C-rates. The first values are the root-mean-squared-error (RMSE) and the second values (in brackets) are the peak error.}
    \label{tab:error_models}    
\end{table}

\begin{table}
    \begin{subtable}{.5\linewidth}
      \centering
        \begin{tabular}{r | c | c | c}
            & 0.5C & 1C & 2C \\ \hline
            25 $\degC$ & $0.44 \pm 0.02$ & $0.43 \pm 0.03$ & $0.47 \pm 0.03$ \\
            10 $\degC$ & $0.44 \pm 0.02$ & $0.44 \pm 0.04$ & $0.47 \pm 0.03$ \\
            0 $\degC$ & $0.44 \pm 0.02$ & $0.43 \pm 0.03$ & $0.47 \pm 0.05$ \\
        \end{tabular}
        \caption{TSPMe}
    \end{subtable}%
    \begin{subtable}{.5\linewidth}
      \centering
        \begin{tabular}{r | c | c | c}
            & 0.5C & 1C & 2C \\ \hline
            25 $\degC$ & $19.03 \pm 0.19$ & $9.25 \pm 0.23$ & $9.02 \pm 0.14$ \\
            10 $\degC$ & $18.81 \pm 0.06$ & $9.01 \pm 0.04$ & $8.44 \pm 0.04$ \\
            0 $\degC$ & $18.64 \pm 0.13$ & $9.09 \pm 0.09$ & $8.39 \pm 0.06$ \\
        \end{tabular}
        \caption{TDFN}
    \end{subtable} 
    \caption{Computational time (in seconds) for the TSPMe and the TDFN model. The times shown here are the solving times only (they do not include the system assembly time). To obtain reliable measurements each simulation has been run 20 times and the values shown here are the mean and standard deviation of the samples.}
    \label{tab:comp_time}
\end{table}

\subsection{Comparison with experimental data}
Now we compare the TSPMe model with experimental data to assess if the model can be used in real applications. The experimental data is for a commercial cell, the LG M50. For the rate tests, cells were placed inside an Espec thermal chamber and, for accuracy, thermocouples were used to record the temperature inside the chamber. Cell temperature was monitored on the cell mid surface and four cells were used for each C-rate and ambient temperature. One of these cells had two additional thermocouples: one on the positive tab and one on the negative tab. A 10 A Digatron battery cycler was used for the tests. For the temperature measurements, K-type thermocouples from RS components were used, which is accurate up to a standard deviation of $\pm 0.75 \%$. Three temperature settings 0 $\degC$, 10 $\degC$ and 25 $\degC$ were used and for each cycle. The cycle consisted of a constant current (CC) charge step of C/3 to a cut-off voltage of 4.2 V, followed by a constant voltage (CV) step at 4.2 V until the charge current dropped to C/20, then a two-hour long rest period, before a constant current discharge step to 2.5 V and a final two-hour long rest before repeating it but for a different discharge C-rate. Different discharge C-rates were explored: C/10, C/2, 1C and 2C, running two cycles for each discharge rate. To reduce the data size, a variable time step combined with a voltage difference condition was used. For C/10 and C/2 discharge rates we used 60 s or 10 mV sampling and for 1C and 2C discharge rates we used 1 s or 10 mV sampling. Data was recorded once one of the two conditions was satisfied. Note that one of the cells cycled at 1C (cell 791) has not been included in the validation data because it gave faulty temperature measurements. However, the data for that cell is available in the online repository. We would like to highlight here that the purpose of this work is not to parameterise the thermal model for a specific cell, but to derive and validate a reduced thermal-electrochemical model. Therefore, in the simulations presented in this section we have used existing parameter sets and calibrated and validated them against the experimental data.

Still, there are some aspects to discuss about the parameters before we continue. We have used as a starting point the parameters shown in Table~\ref{tab:parameter_values_LGM50}, but some of them needed to be adjusted. The parameters related to the electrochemistry are taken from \cite{Chen2020}. These parameters are assumed to not depend on temperature, except for the reaction rates, which follow an Arrhenius relation. Note that in Table~\ref{tab:parameter_values_LGM50} the diffusion coefficient in the particles is assumed constant, however in real-life it depends on lithium concentration and temperature. Therefore, the diffusion coefficient used in the model is taken as an ``effective coefficient'', and adjusted for each experimental set-up (i.e. each combination of C-rate and temperature), so the constant value mimics the nonlinear dynamics for that particular set-up. That is, for each combination of C-rate and temperature, we change the value of $D_\mrn$ until we get a reasonable qualitative agreement with experimental data, leaving $D_\mrp$ unchanged. This assumption is discussed in detail in \cite{Chen2020}. In addition, for each temperature we have adjusted the initial concentration in the positive electrode, so the initial equilibrium potential matches the rest potential before starting the discharge.

The parameters for the thermal model have been taken from \cite{Taheri2013} and then tuned to match the data (in particular they were tuned for 1C at 25 $\degC$) obtaining the values of $\theta = 2.32\E{6}$ $\mathrm{J} \; \mathrm{K}^{-1} \; \mathrm{m}^{-3}$ and $h = 16$ $\mathrm{W} \; \mathrm{m}^{-2} \; \mathrm{K}^{-1}$. Recall that we have neglected the reversible heat generation term $Q_\mathrm{rev}$. However, for the moderate C-rates in the experimental data its contribution should be relatively small. Finally, the value used for both the cell initial temperature and the ambient temperature (which as a modelling assumption we take them to be the same) is taken to be the average value of the final cell temperature of each experiment. This is because the measured temperature of the thermal chamber might not match the equilibrium temperature of the batteries due to experimental limitations in the chamber temperature controller. The parameters that were adjusted for the comparison with experimental data are shown in Table~\ref{tab:tuned_parameters}.

\begin{table}
    \begin{subtable}{.35\linewidth}
      \centering
        \begin{tabular}{r | c | c | c}
            & 0.5C & 1C & 2C \\ \hline
            25 $\degC$ & 0.9 & 2 & 6 \\
            10 $\degC$ & 0.4 & 1 & 3 \\
            0 $\degC$ & 0.22 & 0.55 & 1.5
        \end{tabular}
        \caption{$D_\mrn$ ($10^{-14}$ m$^2$ s$^{-1}$)}
    \end{subtable}%
    \begin{subtable}{.3\linewidth}
      \centering
        \begin{tabular}{r | c }
            & $c_{\mrp,\mathrm{init}}$ \\ \hline
            25 $\degC$ & 17150 \\
            10 $\degC$ & 17750 \\
            0 $\degC$ & 18150
        \end{tabular}
        \caption{$c_{\mrp,\mathrm{init}}$  (mol m$^{-3}$)}
    \end{subtable}%
    \begin{subtable}{.35\linewidth}
      \centering
        \begin{tabular}{r | c | c | c}
            & 0.5C & 1C & 2C \\ \hline
            25 $\degC$ & 24.45 & 24.68 & 24.30 \\
            10 $\degC$ & 9.80 & 10.10 & 9.60 \\
            0 $\degC$ & 0.02 & 0.35 & -0.30 \\
        \end{tabular}
        \caption{$T_\mathrm{amb}$ ($\degC$)}
    \end{subtable}
    \caption{Values for the parameters that have been tuned for the comparison with experimental data.}
    \label{tab:tuned_parameters}    
\end{table}

\begin{figure}[!tb]
    \centering
    \includegraphics[scale=1]{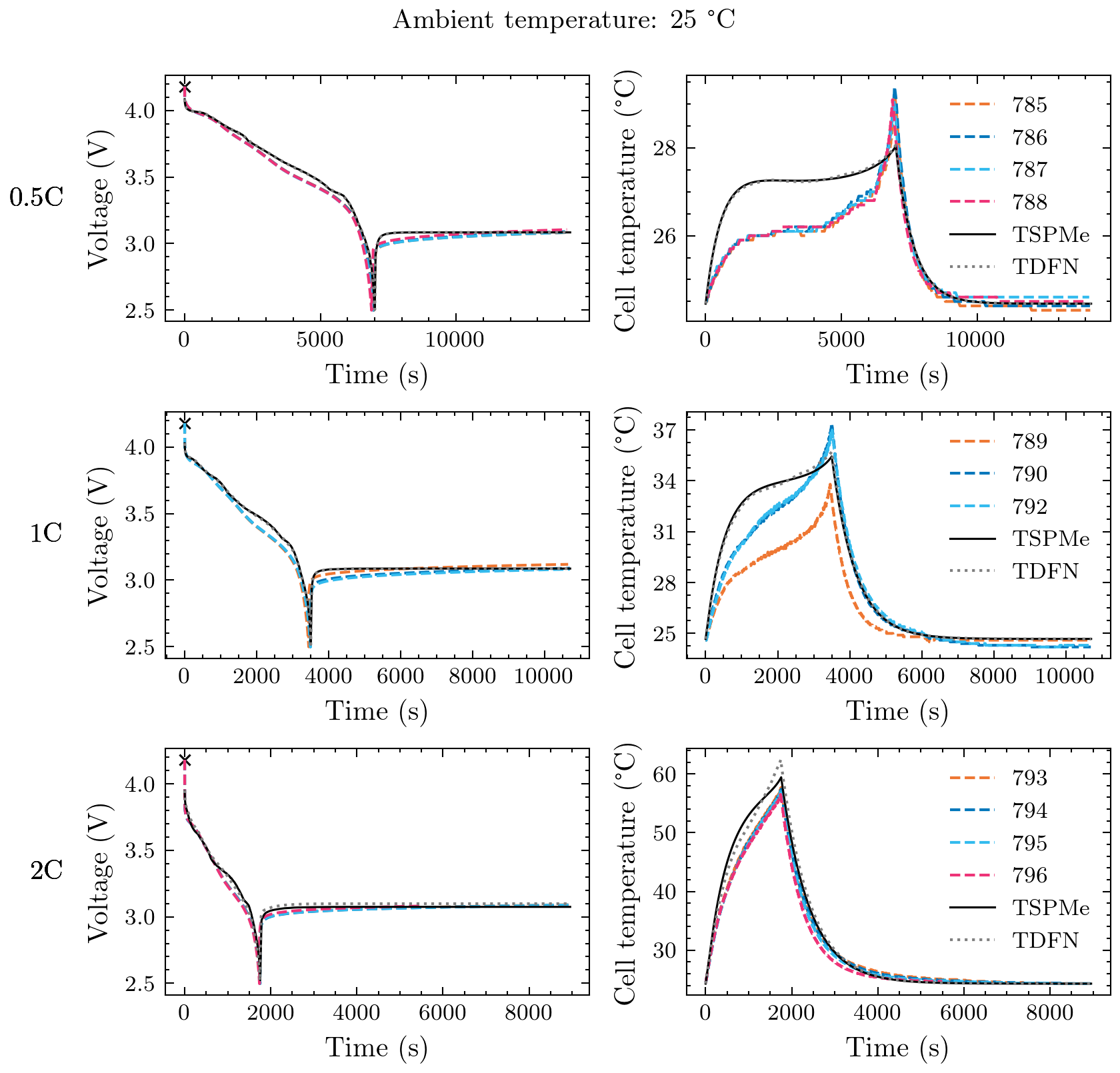}
    \caption{Comparison between TSPMe, TDFN and experimental data at 25 $\degC$. The colour solid lines represent the experimental data for the different cells studied (with the cell number matching that of the dataset), the black solid line represents the TSPMe and the grey dotted line represents the TDFN model. The black cross represents the initial equilibrium potential of the model.}
    \label{fig:comp_exp_25degC}
\end{figure}

\begin{figure}[!htb]
    \centering
    \includegraphics[scale=1]{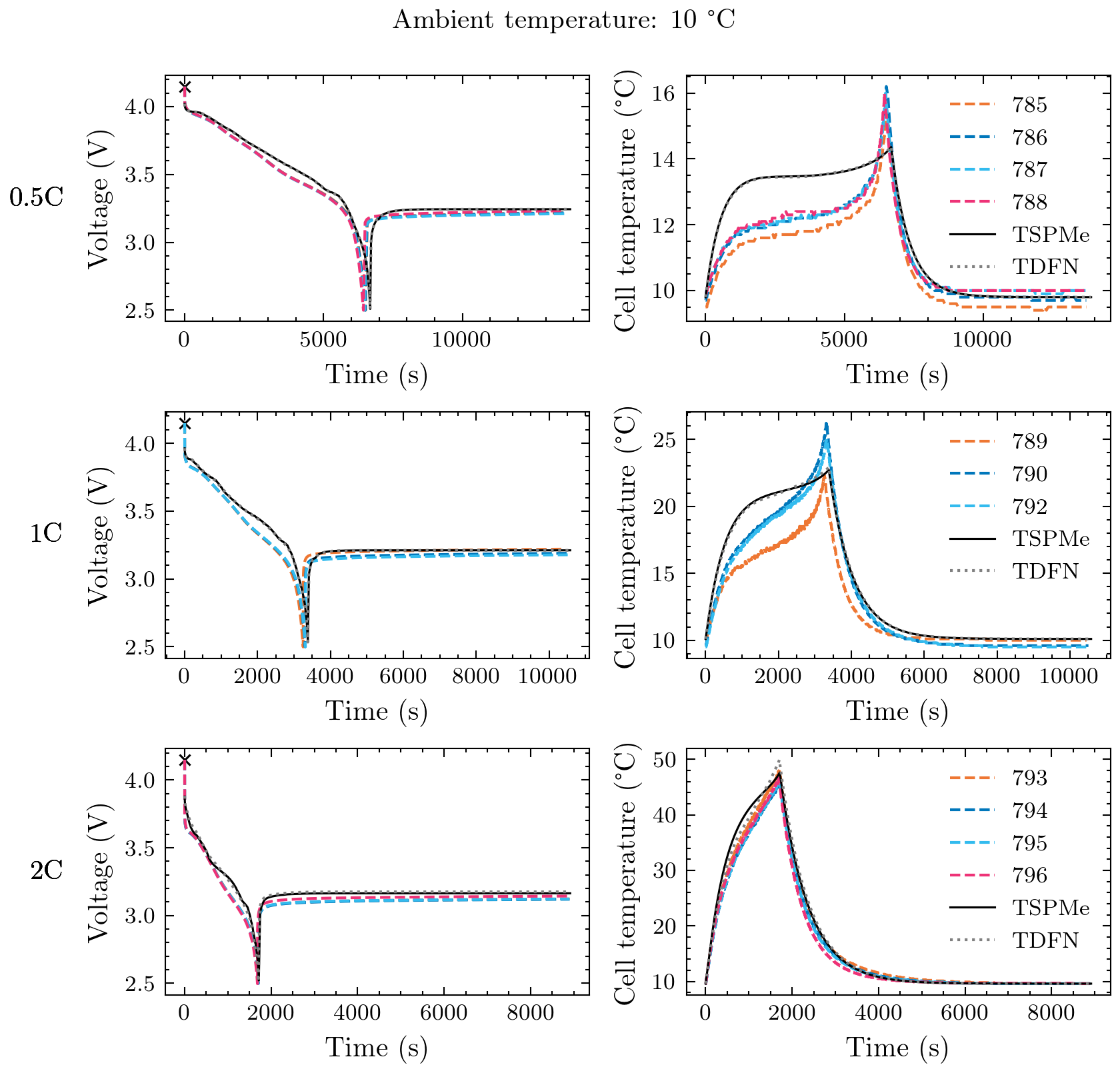}
    \caption{Comparison between TSPMe, TDFN and experimental data at 10 $\degC$. The colour solid lines represent the experimental data for the different cells studied (with the cell number matching that of the dataset), the black solid line represents the TSPMe and the grey dotted line represents the TDFN model. The black cross represents the initial equilibrium potential of the model.}
    \label{fig:comp_exp_10degC}
\end{figure}

\begin{figure}[!htb]
    \centering
    \includegraphics[scale=1]{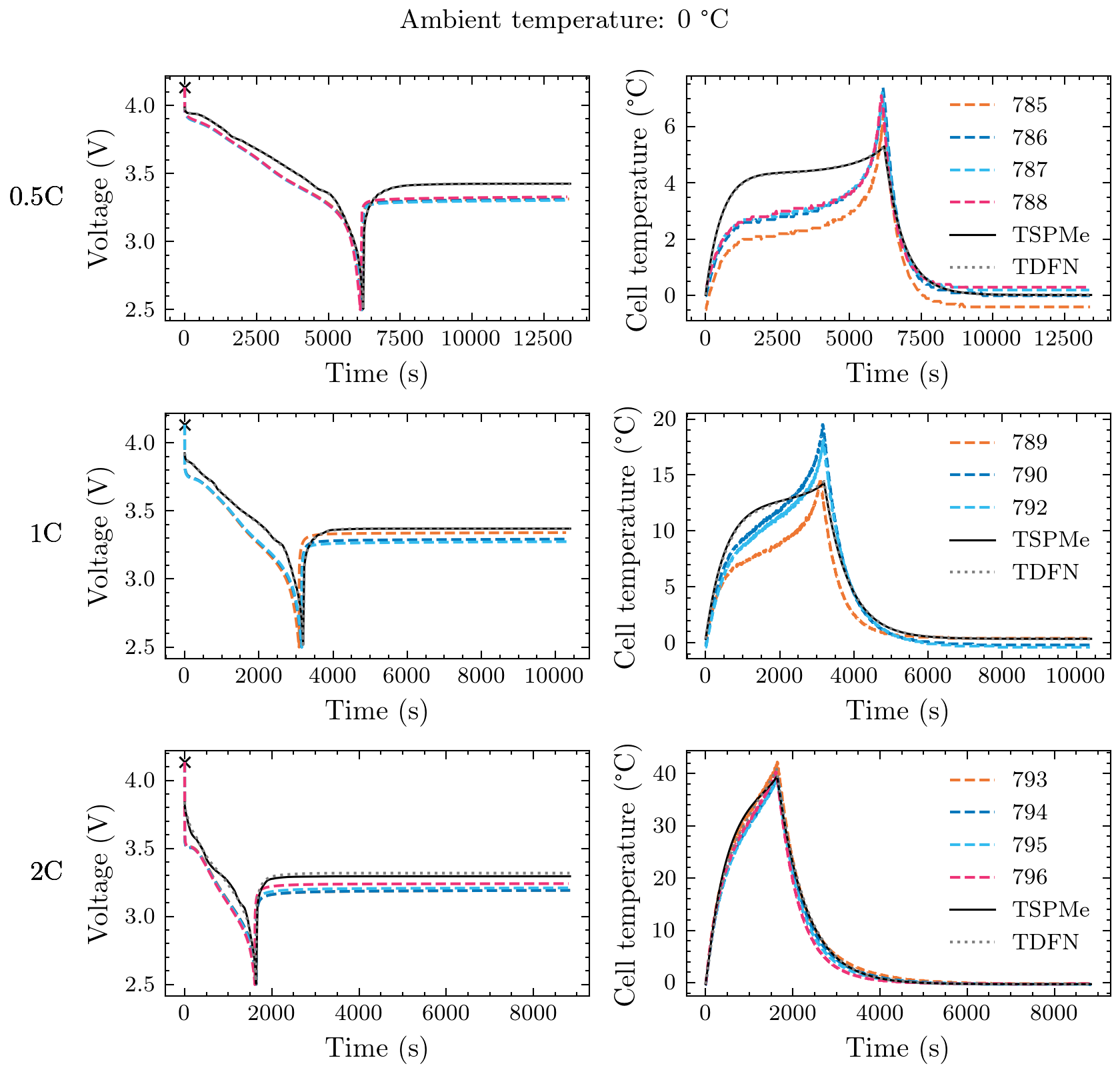}
    \caption{Comparison between TSPMe, TDFN and experimental data at 0 $\degC$. The colour solid lines represent the experimental data for the different cells studied (with the cell number matching that of the dataset), the black solid line represents the TSPMe and the grey dotted line represents the TDFN model. The black cross represents the initial equilibrium potential of the model.}
    \label{fig:comp_exp_0degC}
\end{figure}

The comparison between TSPMe, TDFN and experimental data for voltage and temperature is shown in Figures~\ref{fig:comp_exp_25degC}-\ref{fig:comp_exp_0degC}, with the error metrics for TSPMe shown in Table~\ref{tab:error_data}. The first thing we notice is that, in all cases, the TDFN and TSPMe perform very similarly, therefore the errors with experimental data do not arise from the model reduction process. Looking at Figure~\ref{fig:comp_exp_25degC}, we observe a very good agreement of the voltage at 25 $\degC$ at all C-rates. This could be expected, given that the parameters in \cite{Chen2020} were measured at room temperature. On the other hand, we observe that temperature shows very good agreement with data during the relaxation, which shows that the tuned value of the heat exchange coefficient is a good estimation. The discrepancies observed during the discharge phase, especially at C/2, point out that the discrepancy arises from the heat source terms. It is possible that the issue arises from the irreversible heat source term. If that is the case, given that from \eqref{eq:TSPMe_T} we know that the irreversible heating scales like $i_\mathrm{app} \log (i_\mathrm{app})$ while the heating due to Ohmic losses scale like $i_\mathrm{app}^2$, as the C-rate increases the Ohmic heating dominates and the discrepancy due to the irreversible heat becomes less apparent. The discrepancy could also arise from the reversible heat source term which has not been included in the simulations. In both cases, further work is required to validate these hypotheses.

Looking at the voltage comparison at lower temperatures, we notice that they show worse agreement as the temperature decreases. This is reasonable given that, again, the parameters were measured at room temperature. And even the reaction rates, which have a temperature dependence, were measured in the range between 25 $\degC$ and 60 $\degC$ and therefore the values at 10 $\degC$ and 0 $\degC$ are an extrapolation. Moreover, for temperature, we notice a slight increase in the goodness of fit as the ambient temperature decreases, which is not what we expected. This could also be a side-effect of a discrepancy in the heat source terms, as discussed previously.

\begin{table}
    \begin{subtable}{.55\linewidth}
      \centering
        \begin{tabular}{r | c | c | c}
            & 0.5C & 1C & 2C \\ \hline
            25 $\degC$ & 72.99 (0.97) & 49.94 (0.96) & 35.99 (0.96) \\
            10 $\degC$ & 116.32 (0.89) & 73.89 (0.87) & 61.83 (0.80) \\
            0 $\degC$ & 99.39 (0.91) & 102.41 (0.65) & 100.81 (0.30) \\
        \end{tabular}
        \caption{Voltage (mV)}
    \end{subtable}%
    \begin{subtable}{.45\linewidth}
      \centering
        \begin{tabular}{r | c | c | c}
            & 0.5C & 1C & 2C \\ \hline
            25 $\degC$ & 0.75 (0.67) & 1.46 (0.83) & 2.05 (0.96) \\
            10 $\degC$ & 0.98 (0.67) & 1.50 (0.88) & 1.56 (0.98) \\
            0 $\degC$ & 1.09 (0.72) & 1.46 (0.92) & 1.11 (0.99) \\
        \end{tabular}
        \caption{Temperature ($\degC$)}
    \end{subtable}
    \caption{Error between TSPMe and experimental data for different temperatures and C-rates. The first values are the root-mean-squared-error (RMSE) and the second values (in brackets) are the coefficient of determination $R^2$. Note that the latter is dimensionless and the closer it is to one, the better the fit is.}
    \label{tab:error_data}    
\end{table}

After comparing the TSPMe and TDFN against experimental data we conclude that the TSPMe can provide an accuracy similar to the TDFN but at a much lower computational cost. The discrepancies observed between the models and the data do not arise from the model reduction, as they are observed in the TDFN as well, so they most likely come from the parameter set. It is well-known that an accurate parameter set is required to obtain reliable predictions, and the parameter set from \cite{Chen2020} does not take into account the temperature dependence of the parameters. Therefore, future work is required to obtain a more accurate parameter set for thermal-electrochemical models so we can validate the model against the LG M50 experimental data.

\section{Conclusions}
In this work, we have derived a reduced thermal-electrochemical model that can be used to simulate the behaviour of lithium-ion batteries in a fast and accurate way (compared to the thermal Doyle-Fuller-Newman model). The reduced model, which we refer to as Thermal Single Particle Model with electrolyte (TSPMe), is a SPM-type model (Single Particle Model) but, as opposed to many of the SPM-type models found in the literature, the TSPMe has been derived systematically using asymptotic techniques. This systematic method allows us to determine in advance the range of validity of the model and the accuracy of the reduced model.

Before presenting the TSPMe, we discussed the concept of the Single Particle Model (SPM), both for pure electrochemical models and coupled thermal-electrochemical models (Section \ref{sec:SPM_discussion}). From that analysis we concluded that the fundamental feature of SPM-type models is that they split the model into two steps: the first step involves solving a system of partial differential equations (PDEs) to determine the intercalated lithium and electrolyte ion concentrations, while the second step involves finding any other variable of interest from closed form expressions. This is a crucial distinction because the computationally expensive part when solving the model is solving the PDEs. Moreover, when comparing the different SPM-type models in the literature we find that the main differences between them are in the second step, thus the system of PDEs is the same across most of the models. The same idea naturally extends when a thermal-electrochemical model is considered. The reduction of the thermal DFN model to the TSPMe was based on only two assumptions: the deviations from the open-circuit potential are small and the heat transfer of the battery is limited by the heat exchange with the environment. With only these assumptions, we can already perform the model reduction, as any other assumption usually taken in other reduced models came as a consequence of the assumptions and the governing equations.

In terms of performance, we found that the TSPMe approximates very well the thermal DFN model, especially at low C-rates, but it is significantly faster. At high C-rates (2C) it is over ten times faster while at low C-rates (C/2) the TSPMe is over forty times faster. Moreover, despite all the limitations of the model (due to the assumptions taken to derive it) and the parameters used in the simulations (which were estimated at room temperature), the TSPMe showed good agreement with experimental data on a commercial cell. This good accuracy and low computational time make the TSPMe model (with other models of the same family) a very good candidate for battery design and control.

There are two main lines to explore as possible extensions of this work. On one side, a full thermal parameterisation, including thermal parameters and thermal dependence of electrochemical parameters in the relevant temperature range (0 $\degC$ to 25 $\degC$), is required for a better comparison between the TSPMe and experimental data. This will allow us to rule out discrepancies due to the parameters rather than the model, and thus obtain a more critical validation of the model. On the other side, the model should also be validated against models resolving the full jellyroll structure of the cylindrical battery (e.g. \cite{Tranter2020}) to assess the discrepancies between the TSPMe, which assumes a lumped thermal model, and the fully resolved model.

\section*{Acknowledgements}
This work is supported by the Faraday Institution [EP/S003053/1 grant number FIRG003] and the HVM Catapult project [number 8164]. FBP and WDW would like to thank Prof Emma Kendrick and Kieran O'Regan, from the University of Birmingham, for the useful discussions.

\section*{Data and code availability}
The code and the experimental datasets are available on the repository:

\url{https://github.com/brosaplanella/TEC-reduced-model} (DOI: doi.org/10.5281/zenodo.4085227).

\section*{CRediT author statement}
\textbf{Ferran Brosa Planella:} Conceptualization, Methodology, Software, Formal analysis, Data Curation, Writing – Original Draft. \textbf{Muhammad Sheikh:} Investigation. \textbf{W. Dhammika Widanage:} Conceptualization, Writing – Review \& Editing, Supervision, Funding acquisition.

\bibliographystyle{unsrt}
\bibliography{references}

\appendix

\section{Derivation of the non-dimensional model}
In order to perform the asymptotic analysis we write the full model in dimensionless form. Therefore, we start stating the dimensional version of a  thermal DFN model and proceed to non-dimensionalise it. More details in the derivation of this model can be found in \cite{Plett2015}.

\begin{figure}
    \centering
    \includegraphics[scale=1]{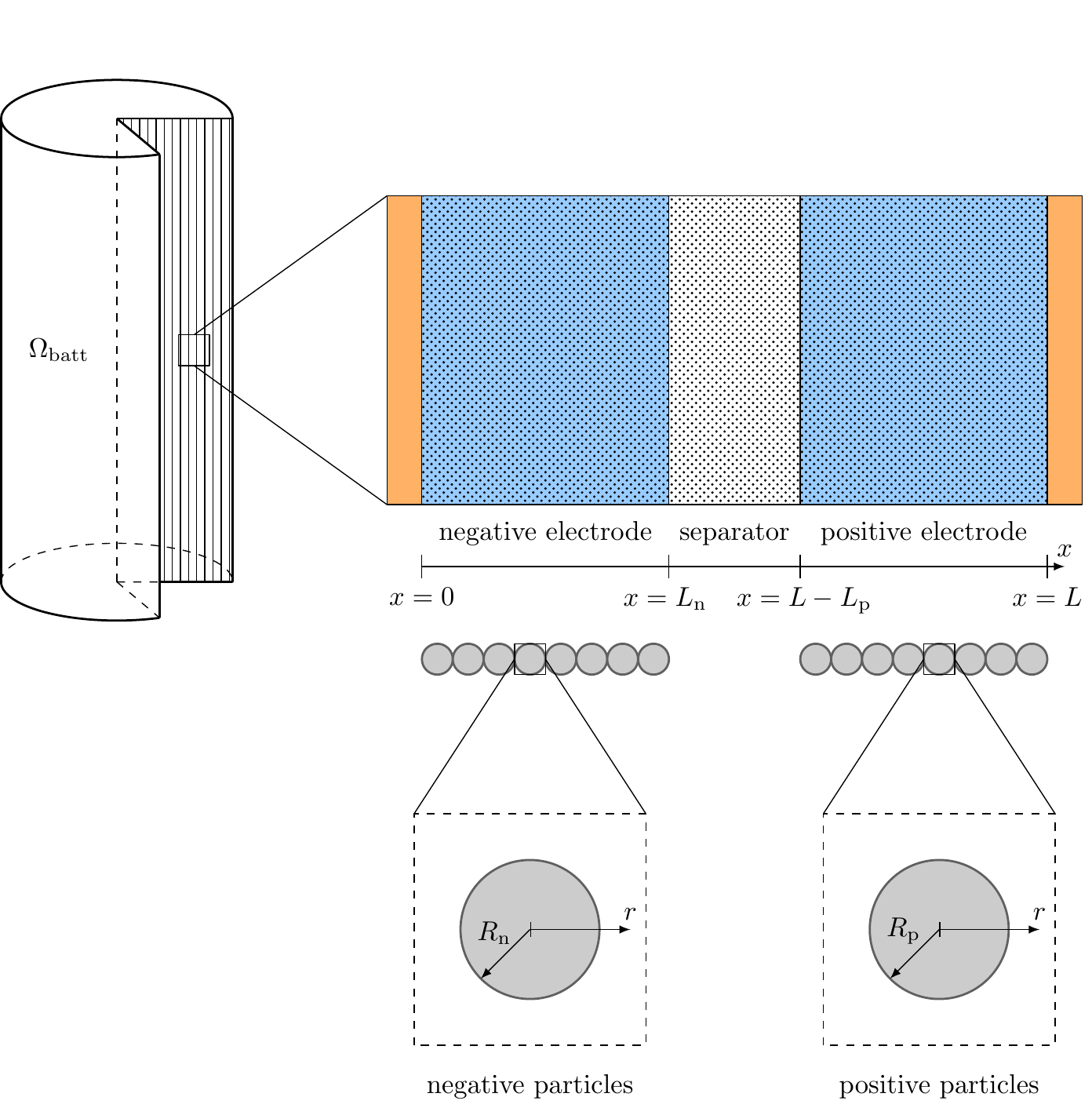}
    \caption{Geometry for the thermal DFN model. The battery is the domain $\mathbf{y} \in \Omega_\mathrm{batt}$ which can have an arbitrary geometry (here, for example, it is depicted as a cylindrical domain). It is composed of multiple cells which are considered to be one-dimensional in the spatial coordinate $x$. In turn, porous electrode is modelled as an array of spherical particles, which are described by a one-dimensional spherically symmetric model in the spatial coordinate $r$.}
    \label{fig:sketch_DFN}
\end{figure}

The geometry of the model is shown in Figure~\ref{fig:sketch_DFN}. There are three scales involved in the problem. The thermal model is posed at the largest scale $\mathbf{y}$, which corresponds to the battery. The battery is defined in the domain $\mathbf{y} \in \Omega_\mathrm{batt}$. In turn, the battery is composed of multiple cells and it is at the cell level, defined by $0 \leq x \leq L$, where we pose the electrochemical model. We choose a DFN model and, therefore, in each electrode we have an array of particles in which lithium diffuses. Then, the particles are defined to have the same size and thus the particle domains are $0 \leq r \leq R_k$ for $k \in \{\mrn,\mrp\}$.

Usually, the DFN model defines the effective properties in the porous media using the Bruggeman correlation \cite{Tjaden2016}. However, here we introduce a generic geometric factor $\mathcal{B}$ that accounts for the variation in the bulk transport properties. In this article, the factor is assumed to be a scalar as we consider a one-dimensional model, but it is related to the tensor $\mathcal{B}$ introduced in \cite{Hunt2020}, which arises from the asymptotic homogenisation of the microscale model. Then, we can choose $\mathcal{B}$ to be the Bruggeman correlation to retrieve the classic DFN model \cite{Fuller1994}, or use other approaches, such as Rayleigh's expressions for lattice-arranged spheres \cite{Rayleigh1892}, numerical computations from tomography imaging \cite{Cooper2016} or direct experimental estimation using impedance spectroscopy \cite{Landesfeind2016}. For a critical discussion of these approaches see \cite{Usseglio-Viretta2018,Nguyen2020}. Our model is based on the standard formulation of the DFN model \cite{Fuller1994,Fuller1994a} and therefore we do not include double-layer capacitance effects in the model, but the same reduction framework would hold if they were included (see \cite{Moyles2019asymptotic} for details).

Then, the electrochemical model reads as follows. In each particle, the lithium concentration is governed by
\begin{subequations}
\begin{align}
\pdv{c_{k}}{t} &= \frac{1}{r^2} \pdv{}{r} \left(r^2 D_{k} \pdv{c_{k}}{r} \right), & \quad \text{ in } 0 < r < R_k,\\
\pdv{c_{k}}{r} &= 0, & \quad \text{ at } r = 0,\\
- D_{k} \pdv{c_{k}}{r} &= \frac{J_k}{a_k F}, & \quad \text{ at } r = R_k,\\
c_{k} &= c_{k,\mathrm{init}}, & \quad \text{ at } t = 0,
\end{align}
and the potential in each electrode is described by
\begin{align}
\pdv{i_{k}}{x} &= -J_k,\\
i_{k} &= -\sigma_{k} \pdv{\Phi_{k}}{x},
\end{align}
In both cases, the $x$ domain is defined to be $0 \leq x \leq L_\mrn$ if $k = \mrn$, and $L - L_\mrp \leq x \leq L$ if $k = \mrp$.

The electrolyte equations, which are defined in the domain $0 \leq x \leq L$, are
\begin{align}
\varepsilon(x) \pdv{c_{\mre}}{t} &= - \pdv{N_{\mre}}{x} + \frac{J}{F},\\
\pdv{i_{\mre}}{x} &= J,
\end{align}
with
\begin{align}
N_{\mre} &= -D_\mre \mathcal{B}(x) \pdv{c_{\mre}}{x} + t^+ \frac{i_{\mre}}{F},\\
i_{\mre} &= -\sigma_{\mre} \mathcal{B}(x) \left(\pdv{\Phi_{\mre}}{x} - 2(1-t^+) \frac{R T}{F} \pdv{\log c_{\mre}}{x} \right).
\end{align}

The intercalation reaction between the electrode and the electrolyte is given by
\begin{align}
J &= \begin{cases}
J_\mrn, & \text{ if } 0 \leq x \leq L_\mrn,\\
0, & \text{ if } L_\mrn < x \leq L - L_\mrp,\\
J_\mrp, & \text{ if } L - L_\mrp < x \leq L,
\end{cases}\\
J_k &= a_k j_k \sinh \left( \frac{1}{2} \frac{F}{R T} \eta_k \right),\\
\eta_k &= \Phi_{k} - \Phi_{\mre} - U_k\left( \left. c_{k} \right|_{r = R_k} \right),\\
j_k &= m_k \left. \sqrt{c_{\mre} c_{k} \left(c_{k}^{\max} - c_{k} \right)} \right|_{r = R_k}.
\end{align}
\end{subequations}

The boundary conditions are the following. At the current collector ends we impose
\begin{subequations}
\begin{align}
i_{\mrn} &= i_\mathrm{app}, & N_{\mre} &= 0, & \Phi_{\mre} &= 0, & \text{ at } x &= 0,\\
i_{\mrp} &= i_\mathrm{app}, & N_{\mre} &= 0, & i_{\mre} &= 0, & \text{ at } x &= L,
\end{align}
and using conservation of charge in the electrolyte we can derive the additional condition $i_{\mre} = 0$ at $x=0$, which can be helpful in some situations.

At the electrode-separator interfaces, we impose zero current in the electrodes
\begin{align}
i_{\mrn} &= 0, & \text{ at } x = L_\mrn,\\
i_{\mrp} &= 0, & \text{ at } x = L - L_\mrp.
\end{align}

Finally, we impose the initial condition for the electrolyte concentration
\begin{align}
c_{\mre} &= c_{\mre,\mathrm{init}}, & \text{ at } t = 0.
\end{align}
\end{subequations}

The heat equation at the macroscale is given by
\begin{subequations}
\begin{align}
\theta \pdv{T}{t} &= \kappa \nabla^2 T + Q, & \text{ in } \mathbf{y} \in \Omega_\mathrm{batt},\\
- \kappa \nabla T \cdot \vb{n} &= h (T - T_\mathrm{amb}), & \text{ at } \mathbf{y} \in \partial \Omega_\mathrm{batt},\\
T &= T_\mathrm{amb}, & \text{ at } t = 0,
\end{align}
with 
\begin{equation}
    Q = -\int_0^L i_k \pdv{\Phi_k}{x} \dd x - \int_0^L i_\mre \pdv{\Phi_\mre}{x} \dd x + \int_0^L J_k \eta_k \dd x + \int_0^L J_k \Pi_k \dd x.
\end{equation}
\end{subequations}
To simplify the notation, when integrating an electrode variable over the domain $0 \leq x \leq L$ we imply splitting the integral, taking for $k = \mrn$ the domain $0 \leq x \leq L_\mrn$ and for $k = \mrp$ the domain $L - L_\mrp \leq x \leq L$.

We allow some of the parameters in the model to depend on certain variables. In particular, we take $D_k$ to depend on the lithium concentration $c_k$, and $D_\mre$ and $\sigma_\mre$ to depend on the ion concentration $c_\mre$. Here, we take $t^+$ to be a constant, which is a common modelling assumption. However, the results in this work can be generalised to account for a variable $t^+$. The rest of the parameters are taken to be constant, unless stated otherwise.

We define the following scalings of the problem
\begin{equation}\label{eq:scalings}
\begin{aligned}
t &= t_0 \hat{t}, & c_{k} &= c_{k}^{\max} \hat{c}_{k}, & \Phi_{k} &= \Phi_0 \hat{\Phi}_{k}, & i_{k} &= i_0 \hat{i}_{k}, & J_k &= \frac{i_0}{L} \hat{J}_k, \\ 
x &= L \hat{x}, & c_\mre &= c_{\mre,\mathrm{init}} \hat{c}_\mre, & \Phi_{\mre} &= \frac{R T_\mathrm{amb}}{F} \hat{\Phi}_{\mre}, & i_{\mre} &= i_0 \hat{i}_{\mre} & j_k &= \frac{i_0}{a_k L} \hat{j}_k, \\
r_k &= R_k \hat{r}_k, & D_k &= D_{k,\typ} \hat{D}_k, & \eta_k &= \frac{R T_\mathrm{amb}}{F} \hat{\eta}_k, & i_\mathrm{app} &= i_0 \hat{i}_\mathrm{app}, & T &= \frac{R T_\mathrm{amb} c_{\mrn}^{\max}}{\theta} \hat{T} + T_\mathrm{amb}, \\
\sigma_\mre &= \sigma_{\mre,\typ} \hat{\sigma}_\mre, & D_\mre &= D_{\mre,\typ} \hat{D}_\mre, & U_k &= \Phi_0 \hat{U}_k, & N_{\mre} &= \frac{D_{\mre,\typ} c_{\mre,\mathrm{init}}}{L} \hat{N}_{\mre}, & Q &= \frac{i_0 R T_\mathrm{amb}}{L F} \hat{Q},\\
\mathbf{y} &= L_\mathrm{batt} \hat{\mathbf{y}}, & \Omega_\mathrm{batt} &= L_\mathrm{batt} \hat \Omega_\mathrm{batt}, & \Pi_k &= \frac{R T_\mathrm{amb} }{F} \hat \Pi_k,
\end{aligned}
\end{equation}
and we choose the time scale $t_0$ to be the discharge time scale
\begin{equation}
    t_0 = \frac{F c_{\mrn}^{\max} L}{i_0}.
\end{equation}

The parameters $i_0$ and $\Phi_0$ are the typical current and electrode potential, respectively, and the subscript typ denotes the typical value of that parameter.

Then, we can write the dimensionless model as follows. The model for concentration in the particles reads
\begin{subequations}\label{eq:EC_nondim}
\begin{align}
\mathcal{C}_k \pdv{\hat c_{k}}{\hat t} &= \frac{1}{\hat r^2} \pdv{}{\hat r} \left(\hat r^2 \hat D_{k} \pdv{\hat c_{k}}{\hat r} \right), & \quad \text{ in } 0 < \hat r < 1,\\
\pdv{\hat c_{k}}{\hat r} &= 0, & \quad \text{ at } \hat r = 0,\\
- \hat D_{k} \pdv{\hat c_{k}}{\hat r} &= \frac{\mathcal{C}_k}{\alpha_k \gamma_k} \hat J_k, & \quad \text{ at } \hat r = 1,\\
\hat c_{k} &= \mu_k, & \quad \text{ at } \hat t = 0,
\end{align}
and the potential in each electrode is given by
\begin{align}
\pdv{\hat i_{k}}{\hat x} &= -\hat J_k,\\
\hat i_{k} &= - \lambda \Sigma_{k} \pdv{\hat \Phi_{k}}{\hat x},
\end{align}
Now, the $\hat x$ domain is defined to be $0 \leq \hat x \leq \ell_\mrn$ if $k = \mrn$, and $1 - \ell_\mrp \leq \hat x \leq 1$ if $k = \mrp$.

The electrolyte equations, which are defined in the domain $0 \leq \hat x \leq 1$, are
\begin{align}
\mathcal{C}_\mre \gamma_\mre \varepsilon(\hat x) \pdv{\hat c_{\mre}}{\hat t} &= - \gamma_\mre \pdv{\hat N_{\mre}}{\hat x} + \mathcal{C}_\mre \hat J,\\
\pdv{\hat i_{\mre}}{\hat x} &= \hat J,
\end{align}
with
\begin{align}
\hat N_{\mre} &= -\hat D_\mre \mathcal{B}(\hat x) \pdv{\hat c_{\mre}}{\hat x} + t^+ \frac{\mathcal{C}_\mre}{\gamma_\mre} \hat i_\mre,\\
\hat i_{\mre} &= -\Sigma_\mre \hat \sigma_{\mre} \mathcal{B}(\hat x) \left(\pdv{\hat \Phi_{\mre}}{\hat x} - 2(1-t^+) (1 + \gamma_T \hat T) \pdv{\log \hat c_{\mre}}{\hat x} \right).
\end{align}

The intercalation reaction between the electrode and the electrolyte is given by
\begin{align}
\hat J &= \begin{cases}
\hat J_\mrn, & \text{ if } 0 \leq \hat x \leq \ell_\mrn,\\
0, & \text{ if } \ell_\mrn < \hat x \leq 1 - \ell_\mrp,\\
\hat J_\mrp, & \text{ if } 1 - \ell_\mrp < \hat x \leq 1,
\end{cases}\\
\hat J_k &= \hat j_k \sinh \left( \frac{1}{2} \frac{\hat \eta_k}{1 + \gamma_T \hat T} \right),\\
\hat \eta_k &= \lambda \left( \hat \Phi_{k} - \hat U_k\left( \left. \hat c_{k} \right|_{\hat r = 1} \right) \right) - \hat \Phi_{\mre},\\
\hat j_k &= \frac{\gamma_k}{\mathcal{C}_{\mathrm{r},k}} \left. \sqrt{\hat c_{\mre} \hat c_{k} \left(1 - \hat c_{k} \right)} \right|_{\hat r = 1}.
\end{align}
\end{subequations}

The boundary conditions at current collector ends are
\begin{subequations}\label{eq:EC_BC_nondim}
\begin{align}
\hat i_{\mrn} &= \hat i_\mathrm{app}, & \hat N_{\mre} &= 0, & \hat \Phi_{\mre} &= 0, & \text{ at } \hat x &= 0,\\
\hat i_{\mrp} &= \hat i_\mathrm{app}, & \hat N_{\mre} &= 0, & \hat i_{\mre} &= 0, & \text{ at } \hat x &= 1,
\end{align}
and at the electrode-separator interfaces are
\begin{align}
\hat i_{\mrn} &= 0, & \text{ at } \hat x = \ell_\mrn,\\
\hat i_{\mrp} &= 0, & \text{ at } \hat x = 1 - \ell_\mrp.
\end{align}
Finally, the initial condition for the electrolyte is
\begin{align}
\hat c_{\mre} &= 1, & \text{ at } \hat t = 0.
\end{align}
\end{subequations}

The dimensionless heat equation reads
\begin{subequations}\label{eq:thermal_nondim}
\begin{align}
\pdv{\hat T}{\hat t} & = \mathcal{K} \hat \nabla^2 \hat T + \hat Q, & \text{ in } \hat{\mathbf{y}} \in \hat \Omega_\mathrm{batt},\\
- \hat \nabla \hat T &= \Bi \hat T, & \text{ at } \hat{\mathbf{y}} \in \partial \hat \Omega_\mathrm{batt},\\
\hat T &= 0, & \text{ at } \hat t = 0.
\end{align}
with
\begin{equation}
\hat Q = -\lambda \int_0^1 \hat i_k \pdv{\hat \Phi_k}{\hat x} \dd \hat x - \int_0^1 \hat i_\mre \pdv{\hat \Phi_\mre}{\hat x} \dd \hat x + \int_0^1 \hat J_k \hat \eta_k  \dd \hat x + \int_0^1 \hat J_k \hat \Pi_k \dd \hat x.
\end{equation}
\end{subequations}

The dimensionless parameters of the model are
\begin{equation}
\begin{aligned}
\mathcal{C}_k &= \frac{R_k^2}{D_{k,\typ} t_0}, & \mathcal{C}_\mre &= \frac{L^2}{D_{\mre,\mathrm{typ}} t_0}, & \mathcal{C}_{\mathrm{r}, k} &= \frac{F}{m_k a_k \sqrt{c_{\mre,\mathrm{init}}} t_0}, & \Sigma_{k} &= \frac{R T_\mathrm{amb}}{F L i_0} \sigma_{\mrs,k}, & \Sigma_{\mre} &= \frac{R T_\mathrm{amb}}{F L i_0} \sigma_{\mre, \mathrm{typ}},\\
\gamma_k &= \frac{c_{k}^{\max}}{c_{\mrn}^{\max}}, & \gamma_\mre &= \frac{c_{\mre,\mathrm{init}}}{c_{\mrn}^{\max}}, & \gamma_T &= \frac{R c_{\mrn}^{\max}}{\theta}, & \mu_k &= \frac{c_{k,\mathrm{init}}}{c_{k}^{\max}}, & \lambda &= \frac{\Phi_0 F}{R T_\mathrm{amb}},\\
\ell_k &= \frac{L_k}{L}, & \alpha_k &= a_k R_k, & \mathcal{K} &= \frac{\kappa t_0}{L_\mathrm{batt}^2 \theta}, & \Bi &= \frac{h L_\mathrm{batt}}{\kappa}.
\end{aligned}
\end{equation}

From \cite{Chen2020,Taheri2013} we find typical values of the dimensionless parameters, which are shown in Table~\ref{tab:ND_parameter_values_LGM50}.

\begin{table}
\centering
\begin{tabular}{| c p{9cm} c c c |}
\hline
\textbf{Symbol} & \textbf{Description} & \textbf{Pos.} & \textbf{Sep.} & \textbf{Neg.} \\ \hline
$\mathcal{C}_k$ & Ratio of solid diffusion to discharge time scales  & $0.60 C$ & - & $9.17\E{-2}C$ \\
$\mathcal{C}_\mre$ & Ratio of electrolyte diffusion to discharge time scales & \multicolumn{3}{c |}{$1.49\E{-2}C$} \\
$\mathcal{C}_{\mathrm{r},k}$ & Radius of active material particles & $0.21C$ & - & $1.08C$ \\
$\Sigma_{k}$ & Ratio of thermal voltage to typical ohmic drop in the solid & $0.55C^{-1}$ & - & $656C^{-1}$ \\
$\Sigma_{\mre}$ & Ratio of thermal voltage to typical ohmic drop in the electrolyte & \multicolumn{3}{c |}{$2.90C^{-1}$} \\
$\gamma_{k}$ & Ratio of maximum lithium concentrations in electrode to maximum concentration in negative electrode & $1.90$ & - & $1$ \\
$\gamma_{\mre}$ & Ratio of typical electrolyte concentration to maximum concentration in negative electrode & \multicolumn{3}{c |}{$3.01\E{-2}$} \\
$\gamma_{T}$ & Ratio of temperature variation to reference temperature & \multicolumn{3}{c |}{$9.67\E{-2}$} \\
$\mu_{k}$ & Initial stoichiometry & $0.2700$ & - & $0.9014$ \\
$\lambda$ & Ratio of electrode thickness to cell thickness & \multicolumn{3}{c |}{$38.94$} \\
$\ell_{k}$ & Ratio of electrode thickness to cell thickness & $0.44$ & $0.07$ & $0.49$ \\
$\alpha_{k}$ & Product of particle radius and surface area density & $2.00$ & - & $2.25$ \\
$\mathcal{K}$ & Ratio of thermal diffusion to discharge time scales & \multicolumn{3}{c |}{$41.8C^{-1}$} \\
$\Bi$ & Ratio of external convection to internal conduction & \multicolumn{3}{c |}{$0.19$} \\ \hline
\end{tabular}
\caption{Dimensionless parameters for the LG M50 cell calculated from the dimensional parameters in Table~\ref{tab:parameter_values_LGM50}. The parameter $C$ is the applied C-rate.}
\label{tab:ND_parameter_values_LGM50}
\end{table}

\section{Derivation of the base TSPMe}\label{sec:derivation_TSPMe}
We now consider the asymptotic limits for the analysis. We take the limits
\begin{equation}
\lambda \gg 1, \quad \quad \mathcal{K} \gg 1, \quad \text{ and } \quad \Bi \ll 1,
\end{equation}
and we assume all the other parameters to be $\order{1}$.

The reasoning behind these limits is the following. The limits $\mathcal{K} \gg 1$ and $\Bi \ll 1$, as shown in \ref{sec:derivation_thermal}, give that the temperature is homogeneous in space, which highly simplifies the temperature contributions in the DFN model. This corresponds to the scenario in which the bottleneck in heat transfer is the heat dissipation to the environment. On the other hand, the limit $\lambda \gg 1$ is the only assumption needed to break the DFN model into an SPMe as, combined with the implicit assumption that $\Sigma_{k}$ and $\Sigma_\mre$ are not small (i.e. $\order{1}$ or larger), implies that any deviations from the equilibrium potential are small. Note that at large C-rates this assumption is no longer true because $\Sigma_k$ and $\Sigma_\mre$ become much smaller than one, but it holds for moderate and low C-rates.

In order to simplify the analysis, we now drop hats from the dimensionless variables as for the rest of the appendix we work with the dimensionless model.

\subsection{Derivation of the reduced thermal model}\label{sec:derivation_thermal}
We start by considering the reduction of the thermal model \eqref{eq:thermal_nondim}, as the results of this analysis are helpful in the reduction of the electrochemical model in \ref{sec:reduction_EC_model}. We use the limits $\lambda \gg 1$, $\mathcal{K} \gg 1$ and $\Bi \ll 1$. For simplicity, we define $\delta = \lambda^{-1}$ as our small parameter, so we scale $\mathcal{K} = \delta^{-1} \tilde{\mathcal{K}}$ and $\Bi = \delta \tilde{\Bi}$. Then, we can introduce these scalings into \eqref{eq:thermal_nondim} which, dropping hats, gives
\begin{subequations}\label{eq:thermal_scaled_delta}
\begin{align}
\delta \pdv{T}{t} & = \tilde{\mathcal{K}} \nabla^2 T + \delta Q, & \text{ in } \mathbf{y} \in \Omega_\mathrm{batt},\\
- \nabla T &= \delta \tilde{\Bi} T, & \text{ at } \mathbf{y} \in \partial \Omega_\mathrm{batt},\\
T &= 0, & \text{ at } t = 0,
\end{align}
with
\begin{equation}
    Q = -\delta^{-1} \int_0^1 i_k \pdv{\Phi_k}{x} \dd x - \int_0^1 i_\mre \pdv{\Phi_\mre}{x} \dd x + \int_0^1 J_k \eta_k \dd x + \int_0^1 J_k \Pi_k \dd x.
\end{equation}
\end{subequations}

We now expand the temperature and the heat source term in powers of $\delta$ as
\begin{subequations}
\begin{align}
    T &= T_0 + \delta T_1 + \delta^2 T_2 + \order{\delta^3},\\
    Q &= Q_0 + \delta Q_1 + \delta^2 Q_2 + \order{\delta^3}.
\end{align}
\end{subequations}

We find that, at leading order, the governing equation is
\begin{subequations}
\begin{align}
\nabla^2 T_0 &= 0,& \quad \text{ in } \mathbf{y}\in \Omega_\mathrm{batt},\\
-\nabla T_0 &= 0,& \quad \text{ at } \mathbf{y} \in \partial \Omega_\mathrm{batt},
\end{align}
\end{subequations}
so we conclude that $T_0 = T_0(t)$. We now look at the $\order{\delta}$ equations to determine $T_0$, which are
\begin{subequations}\label{eq:thermal_O_delta}
\begin{align}
\dv{T_0}{t} & = \tilde{\mathcal{K}} \nabla^2 T_1 + Q_0,& \quad \text{ in } \mathbf{y} \in \Omega_\mathrm{batt},\\
- \nabla T_1 &= \tilde{\Bi} T_0,& \quad \text{ at } \mathbf{y} \in \partial \Omega_\mathrm{batt},\\
T_0 &= 0, & \quad \text{ at } t=0.
\end{align}
\end{subequations}
We can now average (\ref{eq:thermal_O_delta}a) over the whole domain $\Omega_\mathrm{batt}$. Applying the divergence theorem and using (\ref{eq:thermal_O_delta}b) we find
\begin{subequations}\label{eq:lumped_thermal}
\begin{align}
\dv{T_0}{t} & = - a_\mathrm{cool} \tilde{\mathcal{K}} \tilde{\Bi} T_0 + \bar{Q}_0,\\
T_0(0) &= 0,
\end{align}
where
\begin{align}
\bar{Q}_0 &= \frac{1}{\| \Omega_\mathrm{batt} \|} \int_{\Omega_\mathrm{batt}} Q_0 \dd V, &
a_{\mathrm{cool}} &= \frac{\| \partial \Omega_\mathrm{batt} \|}{\| \Omega_\mathrm{batt} \|},
\end{align}
\end{subequations}
which represent, respectively, the battery averaged heat source term and the surface area per unit volume of the battery.

We finally need to write down the leading order source term $Q_0$. If we expand (\ref{eq:thermal_scaled_delta}d) we find that there might be an $\order{\delta^{-1}}$ term given by
\begin{equation}
Q_{-1} = -\int_0^1 i_{k 0} \pdv{\Phi_{k 0 }}{x} \dd x,
\end{equation}
however, as we will see in \ref{sec:reduction_EC_model}, the analysis shows that $\pdv{\Phi_{k 0 }}{x} = 0$ so this term vanishes. Then, we have that the leading order heat source term is
\begin{equation}
Q_0 = -\int_0^1 i_{k 0} \pdv{\Phi_{k 1}}{x} \dd x - \int_0^1 i_{\mre 0} \pdv{\Phi_{\mre 0}}{x} \dd x + \int_0^1 J_{k0} \eta_0 \dd x + \int_0^1 J_{k0} \Pi_{k} (c_{k0}) \dd x.
\end{equation}

After simplifying the electrochemical model in the next section we can provide more detailed expressions for each of the terms that compose $Q_0$.

\subsection{Derivation of the reduced electrochemical model}\label{sec:reduction_EC_model}
We now reduce the electrochemical model to the base SPMe model just considering the limits $\lambda \gg 1$ and using the fact that at leading order the temperature is homogeneous in space, as derived in \ref{sec:derivation_thermal}. We use again $\delta = \lambda^{-1} \ll 1$ as our small parameter and rewrite the dimensionless model accordingly. 

We expand all the variables and derived quantities in powers of $\delta$, using the notation
\begin{subequations}
\begin{align}
\Phi_{k} &= \Phi_{k0} + \delta \Phi_{k1} + \order{\delta^2},\\
i_{k} &= i_{k0} + \delta i_{k1} + \order{\delta^2},
\end{align}
\end{subequations}
and so on.

We expand first (\ref{eq:EC_nondim}f) which at leading order gives
\begin{equation}
-\Sigma_{k} \pdv{\Phi_{k0}}{x} = 0,
\end{equation}
from which we conclude that the potential in the electrodes is spatially homogeneous so $\Phi_{k0} = \Phi_{k0}(t)$. Then, from the leading order term in (\ref{eq:EC_nondim}m) we conclude
\begin{equation}\label{eq:phik_O1}
\Phi_{k0}(t) = U_k \left( \left. c_{k0} \right|_{r = 1} \right).
\end{equation}
Therefore, if the open circuit potential $U_k$ is invertible, we have
\begin{equation}
\left. c_{k0} \right|_{r = 1} = U_k^{-1} \left( \Phi_{k0}(t)  \right),
\end{equation}
and thus the concentration at the boundary of the particles does not depend on $x$, i.e. it is the same for all the particles. Note that, if the open circuit potential $U_k$ is not invertible, then this method no longer works because particles with different concentrations could have the same potential. This is the case of 
lithium iron phosphate (LFP) electrodes which have very flat open circuit potentials \cite{Richardson2020}.

If the concentration is initially homogeneous in $x$, it must remain homogeneous at all times so we conclude that $c_{k0} = c_{k0} (t,r)$. Using (\ref{eq:EC_nondim}c) we conclude as well that $J_{k0} = J_{k0}(t)$. The assumption that the concentration is initially homogeneous in $x$ is reasonable, as that should be the case if the battery is left to rest for long enough.

We now take the leading order expansion of (\ref{eq:EC_nondim}e) with the boundary conditions \eqref{eq:EC_BC_nondim}. Integrating in each electrode separately we find
\begin{equation}\label{eq:J0}
J_{\mrn 0} = \frac{i_\mathrm{app}}{\ell_\mrn} \quad \text{ and } \quad J_{\mrp 0} = -\frac{i_\mathrm{app}}{\ell_\mrp}.
\end{equation}
This is a key result because $J_k$ were the terms coupling the four PDEs together. Now that they can be determined \emph{a priori} the system of PDEs decouples and thus it is a much simpler problem to deal with.

Now we can write write the leading order equations for $c_k$ and $c_\mre$. For the electrode particles we have
\begin{subequations}\label{eq:ck_O1}
\begin{align}
\mathcal{C}_k \pdv{c_{k0}}{t} &= \frac{1}{r^2} \pdv{}{r} \left(r^2 D_{k}(c_{k0}) \pdv{c_{k0}}{r} \right), & \quad \text{ in } 0 < r < 1,\\
\pdv{c_{k0}}{r} &= 0, & \quad \text{ at } r = 0,\\
- D_{k}(c_{k0}) \pdv{c_{k0}}{r} &= \frac{\mathcal{C}_k}{\alpha_k \gamma_k} J_{k0}, & \quad \text{ at } r = 1,\\
c_{k0} &= \mu_k, & \quad \text{ at } t = 0,
\end{align}
\end{subequations}
while for the electrolyte we have
\begin{subequations}\label{eq:ce_O1}
\begin{align}
\mathcal{C}_\mre \gamma_\mre \varepsilon(x) \pdv{c_{\mre 0}}{t} &= \gamma_\mre \pdv{}{x} \left(D_\mre (c_{\mre 0}) \mathcal{B}(x) \pdv{c_{\mre 0}}{x} \right) + (1 - t^+) \mathcal{C}_\mre J_0, & \text{ in } 0 < x < 1,\\
\pdv{c_{\mre 0}}{x} &= 0, & \text{ at } x = 0,1,\\
c_{\mre 0} &= 1, & \text{ at } t = 0,
\end{align}
\end{subequations}
where, to obtain the boundary conditions (\ref{eq:ce_O1}b), we have combined the boundary conditions for molar flux and current in \eqref{eq:EC_BC_nondim}. 

Having the concentration equations that compose the SPMe, we now focus on the potentials. We want to calculate the leading order term of $\Phi_{\mre}$ and the first order term of $\Phi_{k}$ so the potentials, in dimensional form, are accurate up to $\order{\frac{R T_\mathrm{amb}}{F}}$.

We focus first on the leading order electrolyte potential $\Phi_{\mre 0}$. We start calculating $i_{\mre 0}$, and expanding (\ref{eq:EC_nondim}h) jointly with the boundary conditions \eqref{eq:EC_BC_nondim} and integrating we obtain
\begin{equation}
i_{\mre 0} = \begin{cases}
\frac{i_\mathrm{app}}{\ell_\mrn} x, & \text{ for } 0 \leq x \leq \ell_\mrn,\\
i_\mathrm{app}, & \text{ for } \ell_\mrn < x \leq 1 - \ell_\mrp,\\
\frac{i_\mathrm{app}}{\ell_\mrp} (1 - x), & \text{ for } 1 - \ell_\mrp < x \leq 1.
\end{cases}
\end{equation}

Now we can use the leading order expansion of (\ref{eq:EC_nondim}j) to determine the leading order potential. We have
\begin{equation}
i_{\mre 0} = -\Sigma_\mre \sigma_{\mre} \left( c_{\mre 0} \right) \mathcal{B}(x) \left(\pdv{\Phi_{\mre 0}}{x} - 2(1 - t^+) (1 + \gamma_T T_0) \pdv{\log c_{\mre 0}}{x} \right).
\end{equation}
Integrating and using the reference of potential boundary condition in (\ref{eq:EC_BC_nondim}a) we obtain
\begin{equation}
\Phi_{\mre 0} = - \int_0^x \frac{i_{\mre 0}(s)}{\Sigma_\mre \sigma_{\mre} \left(c_{\mre 0}(s) \right) \mathcal{B}(s)} \dd s + 2 (1 - t^+) \log \frac{c_{\mre 0}(x, t)}{c_{\mre 0}(0, t)}.
\end{equation}

Finally, we calculate $\Phi_{k1}$ taking the  $\order{\delta}$ equation of (\ref{eq:EC_nondim}e) combined with the $\order{1}$ term in (\ref{eq:EC_nondim}f), which give
\begin{equation}
-\Sigma_{k} \pdv[2]{\Phi_{k1}}{x} = - J_{k0},
\end{equation}
so we find
\begin{equation}
\Phi_{k1} = \frac{J_{k0}}{2 \Sigma_{k}} x^2 + A_k x + B_k,
\end{equation}
where $A_k$ and $B_k$ are integration constants that need to be determined. These constants are different for each electrode so we determine them separately. Applying the boundary conditions (\ref{eq:EC_BC_nondim}a) and (\ref{eq:EC_BC_nondim}b) we find
\begin{subequations}
\begin{align}
\Phi_{\mrn 1} &= -\frac{i_\mathrm{app} (2 \ell_\mrn - x) x}{2 \ell_\mrn \Sigma_{\mrn}} + B_\mrn,\\
\Phi_{\mrp 1} &= \frac{i_\mathrm{app} (2(1 - \ell_\mrp) - x) x}{2 \ell_\mrp \Sigma_{\mrp}} + B_\mrp.
\end{align}
\end{subequations}
We use (\ref{eq:EC_nondim}l) and (\ref{eq:EC_nondim}m) to determine the values of $B_k$. Combining them we have
\begin{equation}
J_{k0} = j_{k0} \sinh \left( \frac{1}{2 (1 + \gamma_T T_0)} \left(\Phi_{k1} - \Phi_{\mre 0} - \left. c_{k 1} U'_k ( c_{k 0} )\right|_{r = 1} \right) \right),
\end{equation}
which can be transformed into
\begin{equation}\label{eq:inverted_BV}
\Phi_{k1} - \Phi_{\mre0} - \left. c_{k 1} U'_k ( c_{k 0} )\right|_{r = 1} = 2 \arcsinh \left( \frac{J_{k0}}{j_{k0}} \right).
\end{equation}
We could determine $B_k$ directly from the expression above, but that would require calculating $c_{k1}$. Instead, and given that $B_k$ are homogeneous in space, we average \eqref{eq:inverted_BV} over each electrode as this approach does not require calculating $c_{k1}$.

We first show that the averaged $c_{k1}$ over each electrode is zero. Here we show it for the negative electrode only, but the analysis for the positive electrode is analogous. The concentration $c_{\mrn 1}$ follows the problem
\begin{subequations}\label{eq:cs1}
\begin{align}
\mathcal{C}_\mrn \pdv{c_{\mrn 1}}{t} &= \frac{1}{r^2} \pdv{}{r} \left(r^2 \pdv{c_{\mrn 1}}{r}  \right), & \quad \text{ in } 0 < r < 1,\\
\pdv{c_{\mrn 1}}{r} &= 0, & \quad \text{ at } r = 0,\\
- \frac{\alpha_\mrn \gamma_\mrn}{\mathcal{C}_\mrn}\pdv{c_{\mrn 1}}{r} &= J_{\mrn 1}, & \quad \text{ at } r = 1,\\
c_{\mrn 1} &= 0, & \quad \text{ at } t = 0.
\end{align}
\end{subequations}

We now define the $x$-averaged concentration as $\bar{c}_{\mrn 1} = \frac{1}{\ell_\mrn} \int_0^{\ell_\mrn} c_{\mrn 1} \dd x$ and we average \eqref{eq:cs1} over $x$ to obtain the governing equations for $\bar{c}_{\mrn 1}$
\begin{subequations}\label{eq:cs1_bar}
\begin{align}
\mathcal{C}_\mrn \pdv{\bar{c}_{\mrn 1}}{t} &= \frac{1}{r^2} \pdv{}{r} \left(r^2 \pdv{\bar{c}_{\mrn 1}}{r}  \right), & \quad \text{ in } 0 < r < 1,\\
\pdv{\bar{c}_{\mrn 1}}{r} &= 0, & \quad \text{ at } r = 0,\\
- \frac{\alpha_\mrn \gamma_\mrn}{\mathcal{C}_\mrn}\pdv{\bar{c}_{\mrn 1}}{r} &= 0, & \quad \text{ at }r = 1,\\
\bar{c}_{\mrn 1} &= 0, & \quad \text{ at } t = 0.
\end{align}
\end{subequations}
where we have used the fact that $\frac{1}{\ell_\mrn} \int_0^{\ell_\mrn} J_{\mrn 1} \dd x = 0$, which can be shown in the same way that we determined $J_{k0}$ in \eqref{eq:J0}. From \eqref{eq:cs1_bar} we find that $\bar{c}_{\mrn 1} \equiv 0$ and following a similar argument we can show that $\bar{c}_{\mrp 1} \equiv 0$.

We now average \eqref{eq:inverted_BV} over each electrode and isolate $B_\mrn$ and $B_\mrp$, which gives
\begin{subequations}
\begin{multline}
B_\mrn = \frac{i_\mathrm{app} \ell_\mrn}{3 \Sigma_{\mrn}} - \frac{1}{\ell_\mrn \Sigma_\mre} \int_0^{\ell_\mrn} \int_0^x \frac{i_{\mre 0}(s, t) \dd s}{\sigma_{\mre} \left(c_{\mre 0}(s, t) \right) \mathcal{B}(s)} \dd x \\
+ 2 (1 - t^+) (1 + \gamma_T T_0) \frac{1}{\ell_\mrn} \int_0^{\ell_\mrn} \log \frac{c_{\mre 0}(x, t)}{c_{\mre 0}(0, t)} \dd x + \frac{2}{\ell_\mrn} (1 + \gamma_T T_0) \int_0^{\ell_\mrn} \arcsinh \left( \frac{i_\mathrm{app}}{\ell_\mrn j_{\mrn 0}} \right) \dd x,
\end{multline}
\begin{multline}
B_\mrp = - \frac{i_\mathrm{app} (2 \ell_\mrp^2 - 6 \ell_\mrp + 3)}{6 \ell_\mrp \Sigma_{\mrp}} - \frac{1}{\ell_\mrp \Sigma_\mre} \int_{1 - \ell_\mrp}^{1} \int_0^x \frac{i_{\mre 0}(s, t) \dd s}{\sigma_{\mre} \left(c_{\mre 0}(s, t) \right) \mathcal{B}(s)} \dd x \\
+ 2 (1 - t^+) (1 + \gamma_T T_0) \frac{1}{\ell_\mrp} \int_{1 - \ell_\mrp}^{1} \log \frac{c_{\mre 0}(x, t)}{c_{\mre 0}(0, t)} \dd x - \frac{2}{\ell_\mrp} (1 + \gamma_T T_0) \int_{1 - \ell_\mrp}^{1} \arcsinh \left( \frac{i_\mathrm{app}}{\ell_\mrp j_{\mrp 0}} \right) \dd x,
\end{multline}
\end{subequations}
so we can now write the expressions for $\Phi_{\mrn 1}$ and $\Phi_{\mrp 1}$. They are given by
\begin{subequations}\label{eq:phik_Odelta}
\begin{multline}
\Phi_{\mrn 1} = - \frac{i_\mathrm{app} (2 \ell_\mrn - x) x}{2 \ell_\mrn \Sigma_{\mrn}} + \frac{i_\mathrm{app} \ell_\mrn}{3 \Sigma_{\mrn}} - \frac{1}{\ell_\mrn \Sigma_\mre} \int_0^{\ell_\mrn} \int_0^x \frac{i_{\mre 0}(s, t) \dd s}{\sigma_{\mre} \left(c_{\mre 0}(s, t) \right) \mathcal{B}(s)} \dd x \\
+ 2 (1 - t^+) (1 + \gamma_T T_0) \frac{1}{\ell_\mrn} \int_0^{\ell_\mrn} \log \frac{c_{\mre 0}(x, t)}{c_{\mre 0}(0, t)} \dd x + \frac{2}{\ell_\mrn} (1 + \gamma_T T_0) \int_0^{\ell_\mrn} \arcsinh \left( \frac{i_\mathrm{app}}{\ell_\mrn j_{\mrn 0}} \right) \dd x,
\end{multline}
\begin{multline}
\Phi_{\mrp 1} = \frac{i_\mathrm{app} (2 (1 - \ell_\mrp) - x) x}{2 \ell_\mrp \Sigma_{\mrp}} - \frac{i_\mathrm{app} (2 \ell_\mrp^2 - 6 \ell_\mrp + 3)}{6 \ell_\mrp \Sigma_{\mrp}} - \frac{1}{\ell_\mrp \Sigma_\mre} \int_{1 - \ell_\mrp}^{1} \int_0^x \frac{i_{\mre 0}(s, t) \dd s}{\sigma_{\mre} \left(c_{\mre 0}(s, t) \right) \mathcal{B}(s)} \dd x \\
+ 2 (1 - t^+) (1 + \gamma_T T_0) \frac{1}{\ell_\mrp} \int_{1 - \ell_\mrp}^{1} \log \frac{c_{\mre 0}(x, t)}{c_{\mre 0}(0, t)} \dd x - \frac{2}{\ell_\mrp} (1 + \gamma_T T_0) \int_{1 - \ell_\mrp}^{1} \arcsinh \left( \frac{i_\mathrm{app}}{\ell_\mrp j_{\mrp 0}} \right) \dd x.
\end{multline}
\end{subequations}

\subsection{Summary of the TSPMe}\label{sec:summary_TSPMe}
Then, the dimensionless version of the TSPMe is composed of the PDEs \eqref{eq:ck_O1} and \eqref{eq:ce_O1}, with the following expressions for the electrode potentials
\begin{subequations}
\begin{align}
    \Phi_\mrn & \approx \Phi_{\mrn 0} + \lambda^{-1} \Phi_{\mrn 1},\\
    \Phi_\mrp & \approx \Phi_{\mrp 0} + \lambda^{-1} \Phi_{\mrp 1},
\end{align}
\end{subequations}
where the expressions for $\Phi_{k 0}$ and $\Phi_{k 1}$ are given by \eqref{eq:phik_O1} and \eqref{eq:phik_Odelta}. From the potentials we can calculate the terminal voltage of the cell.

The thermal model is given by \eqref{eq:lumped_thermal}. However, with the reduced version of the electrochemical model we can find more explicit expression for the heat source terms. Note that, because the cell problem is the same across the battery domain $\Omega_\mathrm{batt}$, the leading order heat source term $Q_0$ is homogeneous in $\mathbf{y}$ and therefore the volume averaging is trivial.

Substituting the expressions for the potential found in \ref{sec:reduction_EC_model} we can rewrite each contribution to the heat source term as
\begin{subequations}
\begin{align}
Q_{\mrs 0} &= -\int_0^1 i_{k 0} \pdv{\Phi_{k 1}}{x} \dd x = \frac{i_\mathrm{app}^2}{3} \left(\frac{\ell_\mrn}{\Sigma_{\mrn}} + \frac{\ell_\mrp}{\Sigma_{\mrp}} \right),\\
Q_{\mre 0} &= - \int_0^1 i_{\mre 0} \pdv{\Phi_{\mre 0}}{x} \dd x = \int_0^1 \frac{i_{\mre 0}(x)^2}{\Sigma_{\mre} \sigma_{\mre}(c_{\mre 0}) \mathcal{B}(x)} \dd x - 2(1 - t^+) (1 + \gamma_T T_0) \int_0^1 i_{\mre 0}(x) \pdv{\log c_{\mre 0}}{x} \dd x,\\
Q_{\mathrm{irr} 0} &=  \int_0^1 J_{k 0} \eta_{k 0} \dd x = 2 i_\mathrm{app} (1 + \gamma_T T_0) \left( \frac{1}{\ell_\mrn} \int_0^{\ell_\mrn} \arcsinh \left( \frac{i_\mathrm{app}}{\ell_\mrn j_{\mrn 0}} \right) \dd x + \frac{1}{\ell_\mrp} \int_{1 - \ell_\mrp}^{1} \arcsinh \left( \frac{i_\mathrm{app}}{\ell_\mrp j_{\mrp 0}} \right) \dd x \right),\\
Q_{\mathrm{rev}0} &= \int_0^1 J_{k 0} \Pi_{k 0} \dd x = i_\mathrm{app} \left(\Pi_{\mrn 0} - \Pi_{\mrp 0} \right).
\end{align}
\end{subequations}

In particular, we can reduce the second integral in $Q_{\mre 0}$ using integration by parts
\begin{equation}
\begin{aligned}
Q_{\mre 0} &= \int_0^1 \frac{i_{\mre 0}(x)^2}{\Sigma_{\mre} \sigma_{\mre}(c_{\mre 0}) \mathcal{B}(x)} \dd x - 2(1 - t^+) (1 + \gamma_T T_0) \int_0^1 i_{\mre 0}(x) \pdv{\log c_{\mre 0}}{x} \dd x\\
&= \int_0^1 \frac{i_{\mre 0}(x)^2}{\Sigma_{\mre} \sigma_{\mre}(c_{\mre 0}) \mathcal{B}(x)} \dd x - 2(1 - t^+) (1 + \gamma_T T_0) \left( \left. i_{\mre 0} \log c_{\mre 0} \right]_0^1 - \int_0^1 \pdv{i_{\mre 0}}{x} \log c_{\mre 0} \dd x \right)\\
&= \int_0^1 \frac{i_{\mre 0}(x)^2}{\Sigma_{\mre} \sigma_{\mre}(c_{\mre 0}) \mathcal{B}(x)} \dd x + 2(1 - t^+) (1 + \gamma_T T_0) i_\mathrm{app} \left( \frac{1}{\ell_\mrn} \int_0^{\ell_\mrn} \log c_{\mre 0} \dd x - \frac{1}{\ell_\mrp} \int_{1 - \ell_\mrp}^{1} \log c_{\mre 0} \dd x \right).
\end{aligned}
\end{equation}

The dimensional version of these equations is presented in Section \ref{sec:TSPMe}.

\section{Further simplifications}\label{sec:further_simplifications_app}
In this section we provide the derivation for the further simplifications to the model presented in Section \ref{sec:further_simplifications}. The three main simplifications considered are: quasi-steady-state electrolyte concentration, constant electrolyte conductivity and fast lithium diffusion.

\subsection{Quasi-steady state electrolyte concentration}
One simplification for the electrolyte concentration is to take the quasi-steady state problem, which is a valid assumption when the current varies over a larger time scale than the electrolyte diffusion and electrolyte ion generation time scales. This corresponds to the asymptotic limit $\mathcal{C}_\mre = \order{\delta}$ and $\gamma_\mre = \order{\delta}$. In order to obtain analytical solutions to this problem, we also assume that the diffusion coefficient in the electrolyte $D_\mre$ does not depend on the concentration.

Introducing the scalings $\mathcal{C}_{\mre} = \delta \tilde{\mathcal{C}}_\mre$ and $\gamma_\mre = \delta \tilde{\gamma}_\mre$ into (\ref{eq:EC_nondim}g), (\ref{eq:EC_nondim}i) and \eqref{eq:EC_BC_nondim} we find that the problem at leading order is given by
\begin{subequations}\label{eq:ss_electrolyte}
\begin{align}
\pdv{}{x} \left(D_{\mre} \mathcal{B}(x) \pdv{c_{\mre 0}}{x} \right) + \frac{\tilde{\mathcal{C}}_\mre}{\tilde{\gamma}_\mre} (1 - t^+) J_0 &= 0, & \quad \text{ in } 0 < x < 1,\\
\pdv{c_{\mre 0}}{x} &= 0, & \quad \text{ at } x = 0,1.
\end{align}
\end{subequations}
Even though this problem seems well-posed as it has two boundary conditions, from conservation of charge in the electrolyte we find that they are equivalent. Therefore, we need an extra piece of information, which comes from imposing that the total ion concentration must be conserved and can be formally derived from an early time asymptotic analysis. Hence, using the value for the initial concentration, we find
\begin{equation}
\int_0^1 \varepsilon (x) c_{\mre 0} \dd x = \varepsilon_\mrn \ell_\mrn + \varepsilon_\mrs \ell_\mrs + \varepsilon_\mrp \ell_\mrp,
\end{equation}
where $\ell_\mrs = 1 - \ell_\mrp - \ell_\mrn$ is the dimensionless separator thickness.

To integrate \eqref{eq:ss_electrolyte} we split the integral into three domain (negative electrode, separator and positive electrode) so $\mathcal{B}(x)$ and $\varepsilon(x)$ are constants and we use continuity of concentration and flux to match the solutions in each domain, which are given by
\begin{align}
    \left. c_\mre \right]_-^+ &= 0, & \left. \mathcal{B}(x) \pdv{c_\mre}{x} \right]_-^+ &= 0, & \text{ at } x = \ell_\mrn,1-\ell_\mrp,
\end{align}
where $]_-^+$ denotes the difference between the limit from the right and the limit from the left at a given point. Then, we find
\begin{equation}
    c_{\mre 0} = 1 + \frac{\mathcal{C}_\mre i_\mathrm{app} (1 - t^+)}{6 \gamma_\mre D_\mre v_\mathrm{pore}}
    \begin{cases}
    \frac{2 \varepsilon_\mrp \ell_\mrp^2}{\mathcal{B}_\mrp} + \frac{3 \ell_\mrs (2 \varepsilon_\mrp \ell_\mrp + \varepsilon_\mrs  \ell_\mrs)}{\mathcal{B}_\mrs} + \frac{3 \frac{v_\mathrm{pore}}{\ell_\mrn} (\ell_\mrn^2 - x^2) - 2 \varepsilon_\mrn \ell_\mrn^2}{\mathcal{B}_\mrn}, & \text{ if } 0 \leq x < \ell_\mrn,\\
    -\frac{2 \varepsilon_\mrn \ell_\mrn^2}{\mathcal{B}_\mrn} + \frac{2 \varepsilon_\mrp \ell_\mrp^2}{\mathcal{B}_\mrp} + \frac{6 v_\mathrm{pore} (1 - \ell_\mrp - x) - 6 \varepsilon_\mrn \ell_\mrn \ell_\mrs - 3 \varepsilon_\mrs \ell_\mrs^2}{\mathcal{B}_\mrs}, & \text{ if } \ell_\mrn \leq x < 1 - \ell_\mrp,\\
    - \frac{2 \varepsilon_\mrn \ell_\mrn^2}{\mathcal{B}_\mrn} - \frac{3 \ell_\mrs (2 \varepsilon_\mrn \ell_\mrn + \varepsilon_\mrs \ell_\mrs)}{\mathcal{B}_\mrs} + \frac{3 \frac{v_\mathrm{pore}}{\ell_\mrp} \left((1 - \ell_\mrp^2) - (2-x) x \right) + 2 \varepsilon_\mrp \ell_\mrp^2}{\mathcal{B}_\mrp}, & \text{ if } 1- \ell_\mrp \leq x \leq 1.
    \end{cases}
\end{equation}

The dimensional form is presented in \eqref{eq:ce_ss}. An additional simplification to the previous model is to take $\gamma_\mre = \order{1}$ instead, which gives a small variation to the electrolyte concentration. Then, we can take $c_{\mre 0} = 1$.

\subsection{Constant electrolyte conductivity}
Another assumption to simplify the electrolyte potential is to take the electrolyte conductivity $\sigma_{\mre}(c_{\mre 0}) = \sigma_{\mre}$ to be a constant. Then, we can obtain analytical expressions for the integrals involving $\sigma_{\mre}$. 

The first term to consider is the integral arising in the definition of $\Phi_{\mre 0}$. We can use the fact that $\mathcal{B}(x)$ is piecewise constant and split the integral into integrals over each domain. Then we find
\begin{equation}
\frac{1}{\Sigma_{\mre} \sigma_{\mre}} \int_0^x \frac{i_{\mre 0}(s, t)}{\mathcal{B}(s)} \dd s = \frac{1}{\Sigma_{\mre} \sigma_{\mre}}
\begin{cases}
\frac{i_\mathrm{app}}{2 \ell_\mrn \mathcal{B}_\mrn} x^2, & \text{ if } \quad 0 \leq x < \ell_\mrn,\\
\frac{i_\mathrm{app} \ell_\mrn}{2 \mathcal{B}_\mrn} + \frac{i_\mathrm{app}}{\mathcal{B}_\mrs} (x - \ell_\mrn), & \text{ if } \quad \ell_\mrn \leq x < 1 - \ell_\mrp,\\
\frac{i_\mathrm{app} \ell_\mrn}{2 \mathcal{B}_\mrn} + \frac{i_\mathrm{app} \ell_\mrs}{\mathcal{B}_\mrs} + \frac{i_\mathrm{app} \ell_\mrp}{2 \mathcal{B}_\mrp} + \frac{i_\mathrm{app}}{2 \ell_\mrp \mathcal{B}_\mrp} (1 - x)^2, & \text{ if } \quad 1 - \ell_\mrp \leq x \leq 1.\\
\end{cases}
\end{equation}
We can now use these expressions to calculate the integrals in $\Phi_{\mrn 1}$ and $\Phi_{\mrp 1}$, which give
\begin{subequations}
\begin{align}
\frac{1}{\ell_\mrn \Sigma_\mre \sigma_\mre} \int_0^{\ell_\mrn} \int_0^x \frac{i_{\mre 0}(s, t) \dd s}{\mathcal{B}(s)} \dd x &= \frac{i_\mathrm{app} \ell_\mrn}{6 \Sigma_\mre \sigma_\mre},\\
\frac{1}{\ell_\mrp \Sigma_\mre \sigma_\mre} \int_{1 - \ell_\mrp}^{1} \int_0^x \frac{i_{\mre 0}(s, t) \dd s}{\mathcal{B}(s)} \dd x &= \frac{i_\mathrm{app}}{6 \Sigma_\mre \sigma_\mre} \left( \frac{3 \ell_\mrn}{\mathcal{B}_\mrn} + \frac{6 \ell_\mrs}{\mathcal{B}_\mrs} + \frac{2 \ell_\mrp}{\mathcal{B}_\mrp} \right).
\end{align}
\end{subequations}

Finally we need to calculate the integral that appears in $Q_{\mre 0}$, which is
\begin{equation}
\frac{1}{\Sigma_\mre \sigma_\mre}\int_0^1 \frac{i_{\mre 0}(x, t)^2}{\mathcal{B}(x)} \dd x = \frac{i_\mathrm{app}^2}{3 \Sigma_\mre \sigma_\mre} \left( \frac{\ell_\mrn}{\mathcal{B}_\mrn} + \frac{3 \ell_\mrs}{\mathcal{B}_\mrs} + \frac{\ell_\mrp}{\mathcal{B}_\mrp} \right).
\end{equation}

Then, the simplified TSPMe can be written as detailed in \ref{sec:summary_TSPMe} but with the new definitions of potential and heat source term
\begin{subequations}
\begin{multline}
\Phi_{\mre 0} = 2 (1 - t^+) (1 + \gamma_T T_0) \log \frac{c_{\mre 0}(x, t)}{c_{\mre 0}(0, t)} \\
- \frac{1}{\Sigma_{\mre} \sigma_{\mre}}
\begin{cases}
\frac{i_\mathrm{app}}{2 \ell_\mrn \mathcal{B}_\mrn} x^2, & \text{ if } \quad 0 \leq x < \ell_\mrn,\\
\frac{i_\mathrm{app} \ell_\mrn}{2 \mathcal{B}_\mrn} + \frac{i_\mathrm{app}}{\mathcal{B}_\mrs} (x - \ell_\mrn), & \text{ if } \quad \ell_\mrn \leq x < 1 - \ell_\mrp,\\
\frac{i_\mathrm{app} \ell_\mrn}{2 \mathcal{B}_\mrn} + \frac{i_\mathrm{app} \ell_\mrs}{\mathcal{B}_\mrs} + \frac{i_\mathrm{app} \ell_\mrp}{2 \mathcal{B}_\mrp} + \frac{i_\mathrm{app}}{2 \ell_\mrp \mathcal{B}_\mrp} (1 - x)^2, & \text{ if } \quad 1 - \ell_\mrp \leq x \leq 1.\\
\end{cases}
\end{multline}
\begin{multline}
\Phi_{\mrn 1} = - \frac{i_\mathrm{app} (2 \ell_\mrn - x) x}{2 \ell_\mrn \Sigma_{\mrn}} + \frac{i_\mathrm{app} \ell_\mrn}{3 \Sigma_{\mrs, \mrn}} - \frac{i_\mathrm{app} \ell_\mrn}{6 \Sigma_\mre \sigma_\mre} \\
+ 2 (1 - t^+) (1 + \gamma_T T_0) \frac{1}{\ell_\mrn} \int_0^{\ell_\mrn} \log \frac{c_{\mre 0}(x, t)}{c_{\mre 0}(0, t)} \dd x + \frac{2}{\ell_\mrn} (1 + \gamma_T T_0) \int_0^{\ell_\mrn} \arcsinh \left( \frac{i_\mathrm{app}}{\ell_\mrn j_{\mrn 0}} \right) \dd x,
\end{multline}
\begin{multline}
\Phi_{\mrp 1} = \frac{i_\mathrm{app} (2 (1 - \ell_\mrp) - x) x}{2 \ell_\mrp \Sigma_{\mrp}} - \frac{i_\mathrm{app} (2 \ell_\mrp^2 - 6 \ell_\mrp + 3)}{6 \ell_\mrp \Sigma_{\mrs, \mrp}} - \frac{i_\mathrm{app}}{6 \Sigma_\mre \sigma_\mre} \left( \frac{3 \ell_\mrn}{\mathcal{B}_\mrn} + \frac{6 \ell_\mrs}{\mathcal{B}_\mrs} + \frac{2 \ell_\mrp}{\mathcal{B}_\mrp} \right) \\
+ 2 (1 - t^+) (1 + \gamma_T T_0) \frac{1}{\ell_\mrp} \int_{1 - \ell_\mrp}^{1} \log \frac{c_{\mre 0}(x, t)}{c_{\mre 0}(0, t)} \dd x - \frac{2}{\ell_\mrp} (1 + \gamma_T T_0) \int_{1 - \ell_\mrp}^{1} \arcsinh \left( \frac{i_\mathrm{app}}{\ell_\mrp j_{\mrp 0}} \right) \dd x,
\end{multline}
\begin{equation}
Q_{\mre 0} = - 2(1 - t^+) (1 + \gamma_T T_0) i_\mathrm{app} \left( \frac{1}{\ell_\mrp} \int_{1 - \ell_\mrp}^{1} \log c_{\mre 0} \dd x - \frac{1}{\ell_\mrn} \int_0^{\ell_\mrn} \log c_{\mre 0} \dd x \right) + \frac{i_\mathrm{app}^2}{3 \Sigma_\mre \sigma_\mre} \left( \frac{\ell_\mrn}{\mathcal{B}_\mrn} + \frac{3 \ell_\mrs}{\mathcal{B}_\mrs} + \frac{\ell_\mrp}{\mathcal{B}_\mrp} \right).
\end{equation}
\end{subequations}

The dimensional form of these equations is shown in Section \ref{sec:constant_sigmae}. A further simplification to this model is the case where the dimensionless electrolyte conductivity is large (i.e. $\Sigma_\mre \gg 1$). Then, we can simply eliminate the terms with a $\Sigma_{\mre}^{-1}$ factor from $\Phi_{\mre 0}$, $\Phi_{\mrn 1}$, $\Phi_{\mrp 1}$ and $Q_{\mre 0}$ (which are the ones we just calculated).

\subsection{Fast electrode diffusion}
The last simplification is to take the limit $\mathcal{C}_k = \order{\delta}$, which corresponds to fast diffusion in the electrode particles. Then, introduction the scaling $\mathcal{C}_k = \delta \tilde{\mathcal{C}}_k$ into (\ref{eq:EC_nondim}a)-(\ref{eq:EC_nondim}d) and expanding, the leading order equation is given by
\begin{subequations}
\begin{align}
\frac{1}{r^2} \pdv{}{r} \left(r^2 D_{k}(c_{k0}) \pdv{c_{k0}}{r} \right) &= 0, & \quad \text{ in } 0 < r < 1,\\
\pdv{c_{k 0}}{r} &= 0, & \quad \text{ at } r = 0,1,
\end{align}
\end{subequations}
which yields $c_{k0} = c_{k0}(t)$.

At $\order{\delta}$ the problem reads
\begin{subequations}
\begin{align}
\tilde{\mathcal{C}}_k \dv{c_{k0}}{t} &= \frac{1}{r^2} \pdv{}{r} \left(r^2 D_{k}(c_{k0}) \pdv{c_{k1}}{r} \right), & \quad \text{ in } 0 < r < 1,\\
\pdv{c_{k1}}{r} &= 0, & \quad \text{ at } r = 0,\\
- D_{k} (c_{k0}) \pdv{c_{k1}}{r} &= \frac{\tilde{\mathcal{C}}_k}{\alpha_k \gamma_k} J_{k0}, & \quad \text{ at } r = 1,\\
c_{k0} &= \mu_k, & \quad \text{ at } t = 0.
\end{align}
\end{subequations}
Averaging over the domain $0 \leq r \leq 1$ and applying the divergence theorem we conclude that
\begin{subequations}
\begin{align}
    \dv{c_{k0}}{t} &= -\frac{J_{k0}}{\alpha_k \gamma_k},\\
    c_{k0}(t) &= \mu_k,
\end{align}
\end{subequations}
and the dimensional version of this equation is provided in Section \ref{sec:fast_diffusion}.

\section{Open-circuit potentials}\label{sec:OCVs}
\begin{figure}[!tb]
    \centering
    \includegraphics[scale=1]{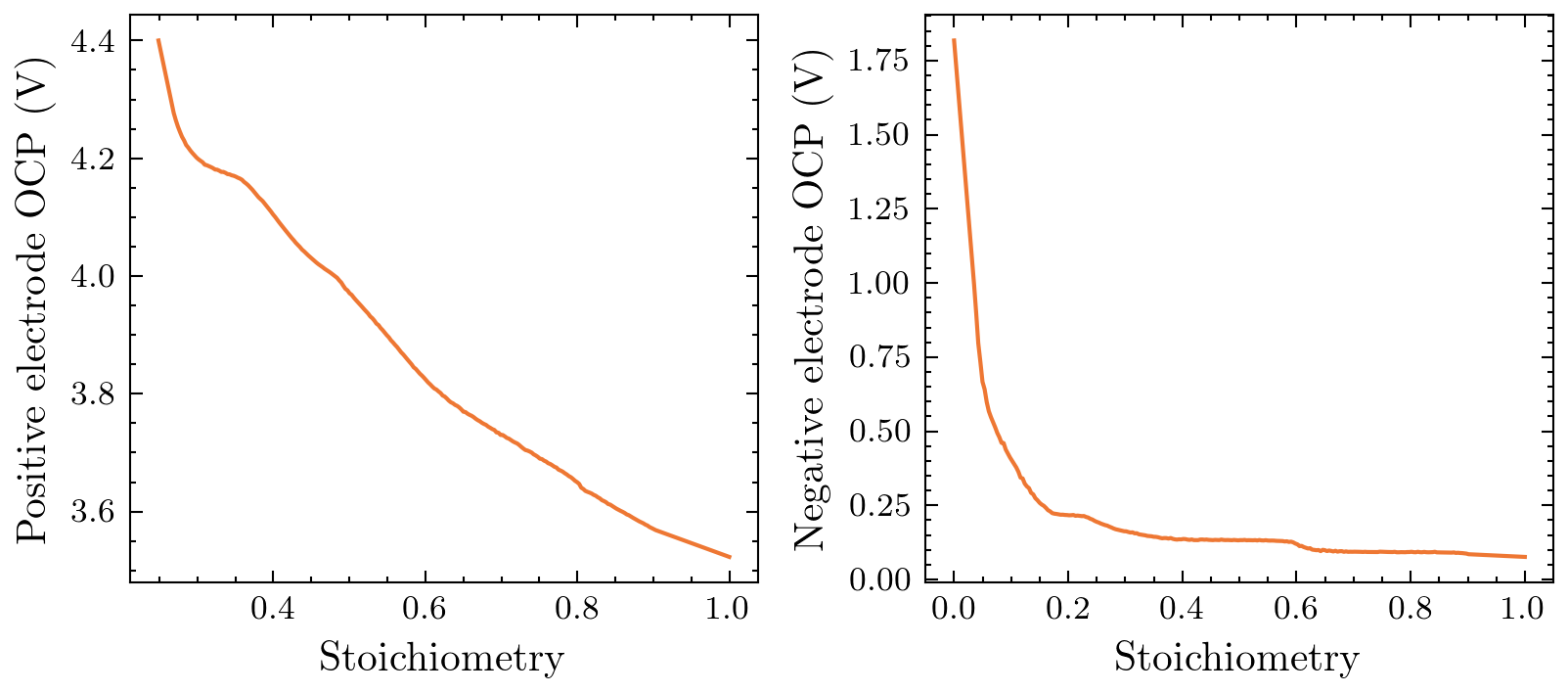}
    \caption{Open-circuit potentials as a function of the electrode stoichiometry for the positive electrode (left) and negative electrode (right). The values have been taken from \cite{Chen2020}.}
    \label{fig:OCVs}
\end{figure}

In the simulations in Section \ref{sec:results} we used the parameter sets from \cite{Chen2020}, and we defined the open-circuit potentials by interpolating the experimental data sets. A plot of the data is shown in  Figure \ref{fig:OCVs}. For the analytical expressions that fit the curves see \cite{Chen2020}.

\end{document}